\documentclass[preprint,authoryear,review,12pt]{elsarticle}

\usepackage{amssymb}
\usepackage{bbold}
\usepackage{amsmath,amssymb,amsthm,bm,mathtools,xspace,booktabs,url}
\usepackage{graphicx,etoolbox}
\usepackage{lscape}

\usepackage{ulem} 

\usepackage{subcaption} 

\journal{arXiv}

\newtheorem{cor}{Corollary}[section]

\newtheorem{prop}{Proposition}[section]

\usepackage{pdfcomment}
\newcommand{\cer}{\mathcal C}
\newcommand{\ZZ}{\mathbb{Z}}

\newcommand{\Vect}[1]{\mathbf{#1}}
 
\newcommand{\cov}{\mathrm{Cov}}
\renewcommand{\cor}{\mathrm{Cor}}
\newcommand{\var}{\mathrm{Var}}
\newcommand{\E}{\mathrm{E}}

\renewcommand{\r}{{\mathcal R}}

\newcommand{\CA}{{\mbox{\scriptsize \sc \,ca}}}
\newcommand{\AC}{{\mbox{\scriptsize \sc \,ac}}}
\newcommand{\lCA}{{\,\ell\mbox{\scriptsize  \sc ca}}}
\newcommand{\R}{{\mbox{\scriptsize \sc \,R}}}
\newcommand{\D}{{\mbox{\scriptsize \sc \,D}}}
\newcommand{\lR}{{\,\ell\mbox{\scriptsize \sc R}}}
\newcommand{\lD}{{\,\ell\mbox{\scriptsize \sc D}}}
\newcommand{\lRD}{{\,\ell\mbox{\scriptsize \sc RD}}}
\newcommand{\RD}{{\,\mbox{\scriptsize \sc RD}}}

\newcommand{\nCA}{{\mbox{\sc ca}}}
\newcommand{\nAC}{{\mbox{\sc ac}}}
\newcommand{\nlCA}{{\ell\mbox{\sc ca}}}
\newcommand{\nR}{{\mbox{\sc r}}}
\newcommand{\nD}{{\mbox{\sc d}}}
\newcommand{\nlR}{{\ell\mbox{\sc r}}}
\newcommand{\nlD}{{\ell\mbox{\sc d}}}
\newcommand{\nlRD}{{\ell\mbox{\sc rd}}}
\newcommand{\nRD}{{\mbox{\sc rd}}}
\newcommand{\SNR}{\mathrm{SNR}}

\usepackage[inline]{trackchanges} 

\usepackage{float}
\makeatletter
\DeclareRobustCommand*\subref{\@ifstar\sf@@subref\sf@subref}
\makeatother

\definecolor{limegreen}{RGB}{50,205,50}

\newcommand{\jf}[1]{{\color{black}{#1}}}

\newcommand{\HL}[1]{{\color{black}{#1}}}

\begin{document} 
	\begin{frontmatter}

		
		
		\title{Inter-regional correlation estimators for functional magnetic resonance imaging}
		
		
		\author[UGACNRSINRIA]{Sophie Achard}
		\author[UGACNRSINRIA]{Jean-Fran\c cois Coeurjolly}
		\author[MONTP,INRIA,IDESP]{Pierre Lafaye de Micheaux}
		\author[UGACNRSINRIA]{Hanâ Lbath}
		\author[CHUVUNIL]{Jonas Richiardi}
		
		\address[UGACNRSINRIA]{Univ. Grenoble Alpes, CNRS, Inria, Grenoble-INP, LJK,
38000 Grenoble, France}
		\address[MONTP]{AMIS, Université Paul Valéry Montpellier 3, France}
		\address[INRIA]{PreMeDICaL - Precision Medicine by Data Integration and Causal Learning, Inria Sophia Antipolis, France}
		\address[IDESP]{Desbrest Institute of Epidemiology and Public Health, Univ Montpellier, INSERM, Montpellier, France}
		\address[CHUVUNIL]{Department of Radiology, Lausanne University Hospital and University of Lausanne, Switzerland}

		\begin{abstract}
			Functional magnetic resonance imaging (fMRI) functional connectivity between brain regions is often computed using parcellations defined by functional or structural atlases. Typically, some kind of voxel averaging is performed to obtain a single temporal correlation estimate per region pair. However, several estimators can be defined for this task, with various assumptions and degrees of robustness to local noise, global noise, and region size.
			
			In this paper, we systematically present and study the properties of 9 different functional connectivity estimators taking into account the spatial structure of fMRI data, based on a simple fMRI data spatial model. These include 3 existing estimators and 6 novel estimators. We demonstrate the empirical properties of the estimators using synthetic, animal, and human data, in terms of graph structure, repeatability and reproducibility, discriminability, dependence on region size, as well as local and global noise robustness.
			
			We prove analytically the link between regional intra-correlation and inter-region correlation, and show that the choice of estimator has a strong influence on  inter-correlation values. Some estimators, including the commonly used \textit{correlation of averages} ($\nCA$), are positively biased, and have more dependence to region size and intra-correlation than robust alternatives, resulting in spatially-dependent bias. We define the new \textit{local correlation of averages} estimator with better theoretical guarantees, lower bias, significantly lower dependence on region size ($p=1.8 e^{-15}$), and significantly higher reproducibility of graph metrics ($p=6.1e^{-5}$) at negligible cost to discriminative power, compared to the $\nCA$ estimator.
			
			The difference in connectivity pattern between the estimators is not distributed randomly throughout the brain, but rather shows a clear ventral-dorsal gradient, suggesting that region size and intra-correlation plays an important role in shaping functional networks defined using the $\nCA$ estimator, and leading to non-trivial differences in their connectivity structure. We provide an open source R package to facilitate the use of the new estimators, together with preprocessed rat time-series.
			
		\end{abstract}
		
		
	\end{frontmatter}
	
	
	\section{Introduction}
	\label{intro}

Functional connectivity of the brain is estimated from observations using non invasive techniques such as electroencephalography (EEG), magnetoencephalography (MEG) or functional Magnetic Resonance Imaging (fMRI). Each recording provides time series associated to spatial locations within regions of the brain. Functional connectomes, that is, graphs representing the estimated connectivity, are then constructed by computing dependence measures between the time series. For fMRI, in addition to the choice of acquisition sequence, physiological noise~\citep{caballero-gaudes_methods_2017}, preprocessing~\citep{braun_testretest_2012}, and subject motion, it has been shown that computation of connectomes is affected by three main parameters: the length of the acquisition~\citep{whitlow_effect_2011,van_dijk_intrinsic_2010}, the number of regions~\citep{stanley_defining_2013,cao_toward_2019} and the chosen frequency band~\citep{cordes_frequencies_2001,salvador_undirected_2005,braun_testretest_2012,chen_bold_2015}. In addition, sample size will also play a role in terms of group comparisons~\citep{termenon2016reliability}.

Typically, each region of the brain, defined by a structural or functional parcellation, is associated to a given set of voxels amongst the thousands for which a signal is  recorded. The idea is then to extract a representative of the set of voxels to attach one time series to each region. When structural atlases are used, the most common approach is to take the average of the voxel time series at each time point. Indeed, almost 70\% of papers on PubMed that used the Human Connectome Project dataset to conduct functional-connectivity-related studies in the last five years (e.g., \cite{Akitoshi2021CA, Figueroa2021CA, Bolt2017CA, Zhang2016CA}) use this method. While there are numerous other approaches to connectivity estimation (including the related techniques for estimating parcellation from connectivity, see, e.g., \cite{eickhoff_connectivitybased_2015}, or using regional medians~\citep{braun_testretest_2012} or eigenvectors~\citep{buchel_modulation_1997,braun_testretest_2012} instead of means), we focus on correlations between averaged regional time-courses, and argue that this technique introduces bias in the estimation of the functional connectomes.\\
    
In the statistical literature, aggregation of data is often used because either the data are collected in different groups (or organizations, or regions), or one wishes to increase the signal to noise ratio. However, difficult challenges arise due to the presence of correlations within the collected datasets. 

\textit{Measurement error} can impact inter-group correlations. For example, in~\cite{ostroff1993comparing}, the objective was to evaluate the relationships among variables by studying correlations between satisfaction and technology at two
	levels: individual scores (i.e., individual correlation) and when individuals are grouped into organizations, also called organizational
	correlation. Depending on three main factors, mainly the measurement errors, the variance within group and the ratio of the intra-group variance and the total variance, 
	the
	ratio of organizational correlation and individual correlation was shown to take values between -1 and~ 2. This means that it is crucial to
	identify the way the data are generated. Measurement errors have been well studied in fMRI. They are related to both the hardware and the subject~\citep{greve_survey_2013}, and are known to impart correlation structure to the data that is not linked to neural activity~\citep{jo_mapping_2010,murphy_resting-state_2013}.

\textit{Group size} can also influence inter-group correlations. In the studies of familial data \citep{rosner.1977.1}, specific characteristics are obtained for
	different families with different sizes.
	The quantification of family resemblance is studied in this context, with the estimation of sibling correlations and/or parent-offspring correlations (see~\cite{donner.1991.1} for a review).  The difficulties arise because of the dependence between the children and the different number of children per family.  It was noted that \textit{correlation between the children and parents} and the
	\textit{average of correlations between all children and parents} are not equal in the majority of cases. In fMRI, the dependence of inter-region correlations on brain region size has been noted~\citep{ssp.2011.1}, and it has been shown that autocorrelation depends on region size~\citep{afyouni_effective_2019}. Regional homogeneity approaches are a wide class of methods where intra-regional connectivity is used to quantify neurobiological differences \citep{jiang_regional_2016}. This observation has also recently been confirmed using Wasserstein distances between intra-correlation densities using Patients with Alzheimer's disease \citep{petersen_quantifying_2016}.

Finally, the \textit{spatial aspect} of the data also complicates correlation estimation. Computing correlations is common when geostatistics is applied to ecology, geography, climate studies, and more. The data collected in these fields are attached to a spatial position and usually with spatial correlation. This problem was first reported by \cite{student.1914.1}, and studied in \cite{clifford.1989.1} for two spatial processes. Applications of these methods can be found for example in the study of meteorological data \citep{gunst.1995.1}, in ecological data \citep{liebhold.1998.1}, but also in fMRI for instance in clustering~ \citep{ye_geostatistical_2009,ye.2011.1}.
	
	In all these fields of application, the main difficulty is to take into account  spatial correlation when the goal is to construct estimators of correlation and to build testing procedures when the averaged variables are not independent.    \\

Thus, in this paper, we question the default choice of using correlations of averages of voxel timecourses, and examine in details the assumptions of various methods and their robustness to various types of noise. 
	We propose first a simple definition of a spatial model of fMRI with intra-correlations within brain regions. Then, computations of correlations are described and we show the potential bias in the estimators. Based on simulations, we illustrate the good behaviour of the newly introduced estimators. Finally, we conclude with results on human data and rat data.    
	
\section{Materials and Methods}

        \subsection{Definition of the proposed spatial model for fMRI data} \label{sec:model}

	Let $\cer\subset\ZZ^d$, $d\in\mathbb{N}^*$, be a finite compact set of multi-indices. In the context of our application, $d=3$ and $\cer$ contains all 3-tuples indexing the voxels of an three-dimensional image of a brain. Each brain is virtually partitioned into $J$ regions of interest which are represented through their subsets of voxels $\r_j$ of cardinal $\#\r_j=N_j$,  $j=1,\ldots,J$. We thus have
	$$
	\cer  = \cup_{j=1}^{J} \r_j \mbox{ and } \# \cer = \sum_{j=1}^J N_j.
	$$
	
For any $d$-tuple $i \in \cer$, a signal $Y_i(\cdot)$ sampled at times $t=1,\ldots,T$ is observed and we assume that it can be decomposed as follows
	\begin{equation} \label{eq:model}
		Y_i(t) = X_i(t) + \varepsilon_i(t) + e(t),
	\end{equation}
where $X_i(\cdot)$ is an unobserved signal of interest, $\varepsilon_i(\cdot)$ is a \textit{local noise} contaminating locally the signal $X_i(\cdot)$, and  $e(\cdot)$ is a \textit{global noise} corrupting in the same way the signal measured in each voxel indexed by an element of $\cer$. This can be, e.g., a consequence of thermal or background noise~\citep{lazar_noise_2008, greve_survey_2013}, which at high field strength has been shown to explain a high proportion of noise variance~\citep{greve_novel_2011}.

	We  make a few assumptions on these different components. First, we assume that all  random variables are centered. We also assume that the signals $X_i(\cdot)$, $\varepsilon_i(\cdot)$ and $e(\cdot)$ are mutually independent and independent in time 
	and that the global noise is homoskedastic with a variance denoted as $\sigma_e\geq0$.
	Note that these are not very strong assumptions, and precise mathematical definitions of these concepts as well as discussions are provided in 
	\ref{appendix.hypothesis}.
	 
	Let us now describe the spatial nature of the signal and of the local noise. For any $i,i^\prime \in \cer$, $ j,j^\prime=1,\ldots, J$ ($j\neq j^\prime$) and for all $t=1,\ldots,T$, we assume that there exists $\sigma_j> 0$, $\sigma_\varepsilon\geq0$, $r_{jj^\prime} \in [-1,1]$, $\rho_{ii^\prime} \in (0,1]$, $\eta_{ii^\prime} \in [-1,1]$ such that  
	\[
	\E[ X_i(t) X_{i^\prime}(t)] = \left\{
	\begin{array}{ll}
		\sigma_j \sigma_{j^\prime} r_{jj^\prime} & \mbox{ if } i \in \r_j, i^\prime\in \r_{j^\prime}, j\neq j^\prime, \\
		\sigma_j^2 \rho_{ii^\prime} & \mbox{ if } i,i^\prime \in \r_j
	\end{array}
	\right.
	\]
	and 
	\[
	\E[ \varepsilon_i(t) \varepsilon_{i^\prime}(t)] = \sigma^2_\varepsilon \eta_{ii^\prime} \mbox{ if } i,i^\prime \in \r_j.
	\]
	The parameter $r_{jj^\prime}$ represents the correlation between two (unobserved) signals of two different regions $\r_j$ and $\r_{j^\prime}$ and is called inter-correlation between regions $\r_j$ and $\r_{j^\prime}$ in the following. \jf{This is the parameter of interest we focus on in the rest of the paper.} The parameter $\rho_{ii^\prime}$ (resp. $\eta_{ii^\prime}$) represents the intra-correlation between two (unobserved) signals (resp. the spatial correlation between two local noises) inside a common region. We assume that inside each region, the signals of interest have positive intra-correlation and that for each time $t$ and for $j=1,\ldots,J$, $\{X_i(t), i \in \r_j\}$ (resp. $\{\varepsilon_i(t), i\in \cer\}$) is a stationary random field defined over $\r_j$ (resp. $\cer$).
	We furthermore assume that both the correlations $\rho_{ii^\prime}$ (for any $i,i^\prime \in \r_j$ for some $j$) and $\eta_{ii^\prime}$ (for $i,i^\prime \in \cer$) \jf{are stationary and isotropic with respect to the uniform distance, that is  $\rho_{ii^\prime}$ and $\eta_{ii^\prime}$} depend only on the (uniform) distance between the two voxels $i$ and $i^\prime$.
	For brevity, we still denote $\rho_{|i^\prime-i|}$ by $\rho_{ii^\prime}$ and $\eta_{|i^\prime-i|}$ by $\eta_{ii^\prime}$ where for $x \in \ZZ^d$, the notation $|x|$ stands for the uniform norm. Our  a priori is that the intra-correlation $\rho_\delta$ is close to~1 for moderate distances $\delta$, meaning that close neighbours are strongly (positively) correlated.
	For the local noise, we assume that there exists $p$ such that the local noise is $p$-dependent, i.e., $\eta_\delta=0$ for any $\delta \ge p$. Without loss of generality, we also intrinsically assume that \textrm{for all} \; $i\in \r_j$ and $i^\prime\in \r_{j^\prime}$, $\varepsilon_i(t)$ and $\varepsilon_{i^\prime}(t)$ are uncorrelated. This simplifies the presentation in the next sections.

	Given a parcellation of the brain, the objective is to estimate inter-correlations $r_{jj^\prime}$ for each pair of regions of interest. We do not model the distribution of $Y_i$ but only its second-order properties (through $X_i, \varepsilon_i, e$). Moreover, we consider the intra-correlations, the correlation of the local noise and the different variances as nuisance parameters that we do not want to estimate. In the next section, we present various estimators of $r_{jj^\prime}$ built in order to address one (or several) of the following problems: (1) the regions of interest $\r_j$ and $\r_{j^\prime}$ may contain a different number of voxels; (2) the intra-correlation may deviate strongly from 1 (especially for large regions); (3) there may be a non negligible local noise $\varepsilon_i$ affecting the signal in each region; (4) there may be a global noise affecting all regions.

	\subsection{Inter-correlation: notation and properties} \label{sec:estimators}

 	Let $\Vect{Y}_1= (Y_1(1),\ldots,Y_1(T))$ and $\Vect{Y}_2= (Y_2(1),\ldots,Y_2(T))$ denote two voxel time-series of length $T$. The notation $\widehat{\cov}(\Vect{Y}_1,\Vect{Y}_2)$, $\widehat{\cor}(\Vect{Y}_1,\Vect{Y}_2)$ and $\widehat{\sigma}^2(\Vect{Y}_1)$ stand for the sample covariance between $\Vect{Y}_1$ and $\Vect{Y}_2$, the  sample correlation between $\Vect{Y}_1$ and $\Vect{Y}_2$ and the sample variance of $\Vect{Y}_1$, respectively. For any $j=1,\ldots,J$, we define a $\nu$-neighbourhood and denote it by $\mathcal V$ as a subset of $n_\nu:=(2\nu+1)^d$  indices, all of which are at a  distance less than or equal to $\nu$ from the center $j$ of the neighbourhood. For any  set of indices $E$ (which could be a $\nu$-neighborhood or a region of interest) and any spatio-temporal field $Z_i(t)$ (which could be $Y_i,X_i,\varepsilon_i$,\dots) we denote by $\bar Z_{E}(t)$ for $t=1,\dots,T$ the time series spatially averaged over $E$, that is
		\[
		\bar Z_{E}(t) = \frac{1}{\# E} \sum_{i\in E} Z_i(t).
		\]
		To sum up, we reserve the bold notation to mainly denote a vector of length~$T$, the notation $\hat{\cdot}$  to denote an average over time while $\bar{\cdot}$ will denote an average over space. Hence, for instance $\hat \sigma^2(\bar{\mathbf{Y}}_{E})$ denote\HL{s} the sample variance of the vector with components $(\# E)^{-1} \sum_{i\in E} Y_i(t)$ for $t=1,\dots,T$. 
		We also let
		\begin{align}
			\bar \rho_{E} = 	\frac1{(\#E)^2} \sum_{i,i^\prime \in E} \rho_{i i^\prime} 
			\qquad \text{and} \qquad
			\bar \eta_{E} = 	\frac1{(\#E)^2} \sum_{i,i^\prime \in E} \eta_{i i^\prime}\HL{.}
			\label{eq:defrhoetaagg} 
		\end{align}
		The quantity $\bar \rho_E$ represents the (spatial) average intra-correlation inside the set $E$. If $E$ corresponds to a $\nu$-neighborhood with moderate $\nu$, we may expect  $\bar \rho_{\mathcal V}$ to be close to 1. Such an observation is probably less realistic when $E=\r_j$ especially for large regions. The quantity $\bar \eta_{E}$ is related to the (spatial) correlation structure of the local noise. By assuming this noise to be $p$-dependent (that is $\eta_\delta=0$ when $\delta \ge p$), it is clear that the larger $\#E$ the smaller $\bar \eta_E$.

	Using the assumption given in Section~\ref{sec:model}, for any $E\subseteq \r_j$ and $E^\prime\subseteq \r_{j^\prime}$, we deduce
			\begin{equation*}
				\cov[\bar Y_E(t) , \bar Y_{E^\prime}(t)] =	\left\{
				\begin{array}{ll}
					\sigma_j \sigma_{j^\prime} r_{jj^\prime} + \sigma^2_e, & \mbox{ if } j\neq j^\prime, \\
					\sigma_j^2 \bar \rho_{E,E^\prime} + \sigma^2_e, & \mbox{ if } j=j^\prime,
				\end{array}
				\right.
			\end{equation*} 
			where 
			\begin{equation*}
				\bar \rho_{E,E^\prime} = \frac{1}{(\#E)(\#E^\prime)} \sum_{i\in E,i^\prime \in E^\prime} \rho_{|i-i^\prime|}.	
			\end{equation*}
		The variance can also be deduced as follows:
	\begin{equation*}
	    \var[ \bar Y_E(t) ] =  \sigma^2_j \; \bar \rho_E + \sigma^2_\varepsilon \; \bar \eta_E  + \sigma^2_e. 
	\end{equation*}
		
	The detail of this result is given in Proposition~\ref{prop:results}.

	To lighten the expression of estimators proposed in the next sections, we define for $j,j^\prime \in \{1,\dots,J\}$
		\begin{equation}\label{eq:morenotation}
			\sigma_{\varepsilon,j} = \frac{\sigma_\varepsilon}{\sigma_j}, \qquad 
			\sigma_{e,j} = \frac{\sigma_e}{\sigma_j}, \qquad 
			\text{ and } \qquad 
			\sigma_{e,jj^\prime} = \frac{\sigma_e}{\sqrt{\sigma_j\sigma_{j^\prime}}}.
		\end{equation}
		In the next sections, we set $j,j^\prime$ and thus aim to estimate $r_{jj^\prime}$.

	
		\begin{figure}[htbp]
		\centering  
		\includegraphics[width=0.8\linewidth]{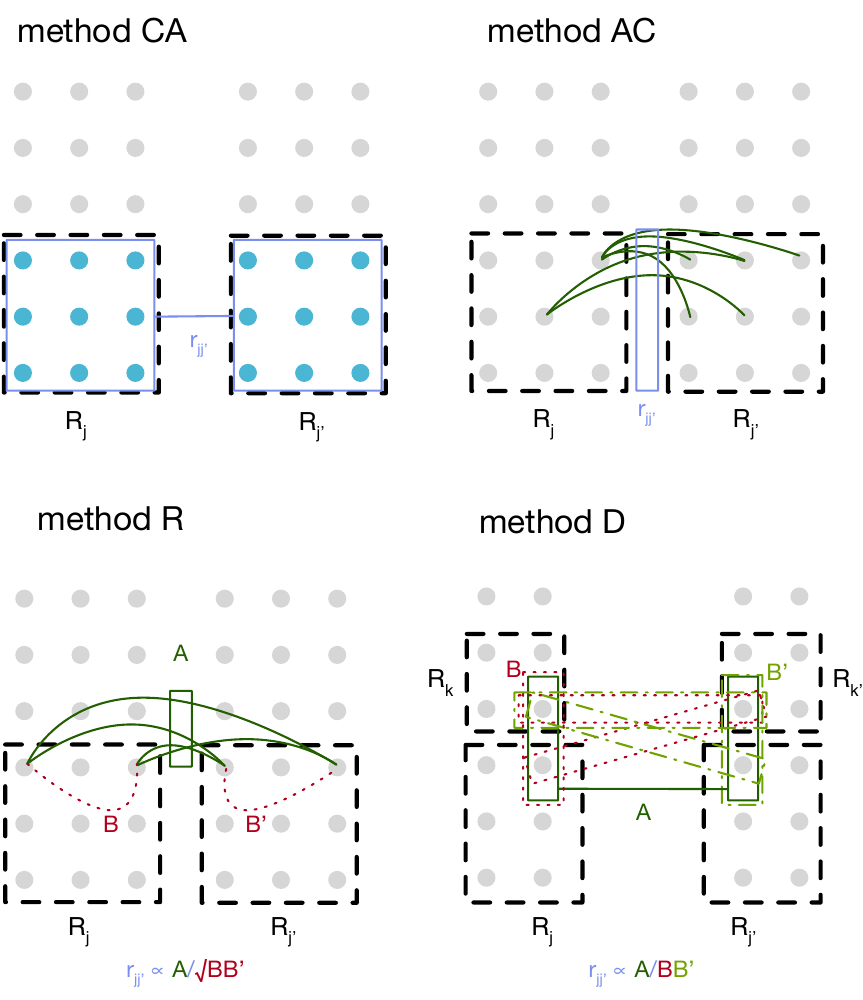}
		\caption{Graphical overview of the main inter-regional correlation estimators discussed in this paper. Gray dots represents voxels. Dashed black lines represent brain regions. Edges between voxels represent voxel-voxel temporal correlations. Blue rectangles show the region and level of aggregation (voxels or correlations). $r_{jj'}$ shows quantities involved in the computation of the inter-regional correlations. Illustrations are approximate, please refer to the relevant equations for the exact definition. Neighbourhood versions of the estimators (starting with $\ell$) use the same principles but involve aggregating in small neighbourhoods within regions.}
		\label{fig:illustrationEstimators}
	\end{figure}

	\subsection{Existing inter-correlation estimators}
	
	We first review existing inter-regional correlation estimators using a unified notation throughout\protect\footnote{In a previous study \citep{ssp.2011.1}, we already described three of the estimators discussed here ($\nCA$, $\nAC$, $\nlCA$), but not with a unified notation, as well as a fourth estimator which is only discussed in the appendix of the present paper}. Results on consistency of the estimators are provided in~\ref{app:proposition1}-\ref{app:consistencylRD}.
	
	\subsubsection{Correlation of averages (method $\nCA$)}
	\label{sec:CA}
	In order to increase the signal-to-noise ratio, \jf{the most standard method in fMRI is to} average (or sometimes convolve with a Gaussian kernel) the signal in space (in each region of interest). The aggregated correlation estimator corresponds to the standard estimator (see Section \ref{intro}) considered for example in \cite{achard.2006.1}, \cite{Bolt2017CA} or \cite{Akitoshi2021CA}:
	\jf{\begin{equation}
			\widehat{r}_{jj^\prime}^{\CA}=\frac{\widehat{\cov}(\Vect{\bar{Y}}_{\r_j},\Vect{\bar{Y}}_{\r_{j^\prime}})}{{\widehat{\sigma}(\Vect{\bar{Y}}_{\r_j})} {\widehat{\sigma}(\Vect{\bar{Y}}_{\r_{j^\prime}})}}.
			\label{eq.agg}
		\end{equation}
	    
	    This estimator, illustrated in Figure~\ref{fig:illustrationEstimators} was designed to reduce the local noise. Indeed, in the absence of global noise ($\sigma_e^2=0$) and assuming equal variances, this estimator tends to $r_{jj^\prime}/\sqrt{(\bar \rho_{\r_j}+\sigma_{\varepsilon}^2 \bar\eta_{\r_j})(\bar \rho_{\r_j^\prime}+\sigma_{\varepsilon}^2 \bar\eta_{\r_j^\prime})}$. The interest of averaging before computing correlations is clear: the local noise is smoothed, thus $\bar \eta_{\r_j}= \mathcal O(1/N_j)$ is probably very small. However, even in absence of noise ($\sigma_\varepsilon=\sigma_e=0$), $\widehat r^\CA$ has a serious drawback since it estimates $r_{jj^\prime}/\sqrt{\bar \rho_{\r_j}\bar \rho_{\r_{j^\prime}}}$ which is highly dependent on intra-correlation. Just to give an example, assume $r_{jj^\prime}=1/2$, $N_j=N_{j^\prime}=2$, $\rho_1=0$ then $\bar \rho_{\r_j} =\bar\rho_{\r_{j^\prime}}=1/2$ and $\widehat r^{\CA}_{jj^\prime}=1$. This is a caricature but illustrates what may happen for  large regions when some of the signals $X_i$ are not enough positively correlated. That fact was already pointed out by~\cite{ssp.2011.1}.}

	\subsubsection{Average of correlations (method \nAC)}
	
	Instead of evaluating correlation of spatial averages, it is natural to perform the (spatial) average of correlations. This estimator, illustrated in Figure~\ref{fig:illustrationEstimators}, is given by:
	\begin{equation}
		\widehat{r}_{jj^\prime}^{\AC}=\frac{1}{N_jN_{j'}}\sum_{i \in \r_j, \\ i' \in \r_{j'}}\widehat{\cor}(\Vect{Y}_i,\Vect{Y}_{i'}).
		\label{eq.ave}
	\end{equation}
	As seen from Table~\ref{tab:summary}, in absence of global noise ($\sigma^2_e=0$) and when the variances are equal to 1, this estimator estimates the quantity $r_{jj^\prime}/(1+\sigma_{\varepsilon}^2)$ which makes this estimator robust to large regions (for which $\bar \rho_{\r_j}$ may be far from 1) but more sensitive to local noise than the estimator $\widehat{r}_{jj^\prime}^{\CA}$.

	\subsubsection{Replicates for correlations (method $\nR$)} \label{sec:R}
	
	In order to \jf{cancel out} the effect of local noise, we introduce to fMRI a slight adaptation of the estimator introduced by~\cite{bergholm.2010.1}, in the context of image analysis. This is based on the concept of replicates within the same region, and denoted by $\widehat{r}^\R$ ($\nR$ for replicates).  It is illustrated in Figure~\ref{fig:illustrationEstimators} and defined by
	\begin{equation}
			\widehat{r}_{jj^\prime}^{\R}=\frac{1}{B} \sum_{b=1}^B \frac{\frac{1}{4}\sum_{\alpha,\beta=1}^2\widehat{\cor}(\Vect{Y}_{i_\alpha^{(b)}},\Vect{Y}_{{i^\prime}_\beta^{(b)}})}{\sqrt{|\widehat{\cor}(\Vect{Y}_{i_1^{(b)}},\Vect{Y}_{i_2^{(b)}}) \; \widehat{\cor}(\Vect{Y}_{{i^\prime}_1^{(b)}},\Vect{Y}_{{i^\prime}_2^{(b)}})}|}
			\label{eq.rave}
	\end{equation}
	where for $b=1,\ldots,B$, $i_1^{(b)},i_2^{(b)} \in\r_j$ are such that $|i_2^{(b)}-i_1^{(b)}|=\delta\ge p$. In the same way, ${i^\prime}_1^{(b)}, {i^\prime}_2^{(b)} \in \r_{j^\prime}$ are such that $|{i^\prime}_2^{(b)}- {i^\prime}_1^{(b)}|=\delta\ge p$. Under equal variances and absence of global noise, $\widehat{r}_{jj^\prime}^{\R}$ estimates $1/|\rho_\delta|$ which is clearly independent of $\sigma_\varepsilon^2$ and may be expected to be close to one if $\delta$ is small.

	
	\subsection{Novel estimators: discarding the effect of global and/or local noise} 
	
	\subsubsection{Use of a priori uncorrelated regions (method $\nD$ based on differences)} 
	\label{sec:rD}

	We now present an estimator which handles the problem of global noise. \jf{To achieve this task, we assume that among the regions where the signal is recorded there are at least two regions say $\r_k$ and $\r_{k^\prime}$ which are uncorrelated between themselves and from all the other ones. With a slight abuse of notation, $k,k^\prime$ will be used for the indices of these two regions, while $j,j^\prime$ will be used when we are interested in the inter-correlation between regions $\r_j$ and $\r_{j^\prime}$ (hence $r_{jk}=r_{jk^\prime}=r_{j^\prime k}=r_{j^\prime k^\prime}=0$).} This assumption is realistic in the context of fMRI data where we are interested in the correlations between cortical regions. Indeed, the field of view is typically larger than the brain itself, and the definition of extra regions is possible, for instance using air voxels or muscle voxels. The estimator is illustrated in Figure~\ref{fig:illustrationEstimators}.
	
	\jf{We propose the following strategy}: for $b=1,\ldots,B$ let $i^{(b)}$, ${i^\prime}^{(b)}$, $k^{(b)}$ and ${k^\prime}^{(b)}$  be voxels of $\r_j$, $\r_{j^\prime}$, $\r_k$ and $\r_{k^\prime}$.
	\begin{equation} \label{eq.save}
		\jf{\widehat{r}^\D_{jj^\prime}} = \frac{1}B \sum_{b=1}^B \widetilde{\cor} ( \Vect{Y}_{i^{(b)}}, \Vect{Y}_{{i^\prime}^{(b)}} ;\Vect{Y}_{k^{(b)}},\Vect{Y}_{{k^\prime}^{(b)}} ),
	\end{equation}
	where for four vectors $\Vect{Y}_1$, $\Vect{Y}_2$, $\Vect{Y}_3$ and $\Vect{Y}_4$ (with same length)
	\begin{equation} \label{eq.cortilde}
		\widetilde{\cor} ( \Vect{Y}_1, \Vect{Y}_2 ;\Vect{Y}_3,\Vect{Y}_4 ) = 
		\frac{\widehat{\cov}(\Vect{Y}_{1} - \Vect{Y}_{3} , \Vect{Y}_{2} - \Vect{Y}_{4} )   }{ \widehat{s} (\Vect{Y}_{1},\Vect{Y}_{3},\Vect{Y}_{4} ) \; \widehat{s} (\Vect{Y}_{2},\Vect{Y}_{3},\Vect{Y}_{4} ) }
	\end{equation}
	and where for three vectors $\Vect{U}$, $\Vect{V}$ and $\Vect{W}$ with same length
	\[
	\widehat{s}^2(\Vect{U},\Vect{V},\Vect{W}) = \left( \; \widehat{\sigma}^2(\Vect{U}-\Vect{V}) + \widehat{\sigma}^2(\Vect{U}-\Vect{W}) - \widehat{\sigma}^2(\Vect{V}-\Vect{W}) \; \right) /2.
	\]
	The intuition of this estimator is quite simple. Assume that the local noise has \HL{zero variance}. Since the noise $e(\cdot)$ is global, \HL{subtracting} from $Y_{i^{(b)}}(t)$ the value $Y_{k^{(b)}}(t)$ and  from $Y_{{i^\prime}^{(b)}}(t)$ the value $Y_{{k^\prime}^{(b)}}(t)$ discards the global noise. And since the regions $\r_k$ and $\r_{k^\prime}$ are not correlated and not correlated to the other ones, the numerator (for each $b$) is an estimate of $\sigma_j \sigma_{j^\prime} r_{jj^\prime}$. Then, we just have to divide by estimates of $\sigma_j$ (and $\sigma_{j^\prime}$). We observe that this cannot be done using simply $\widehat{\sigma}^2(\Vect{Y}_{i^{(b)}} - \Vect{Y}_{k^{(b)}})$ which estimates \jf{$\sigma_j^2+\sigma^2_k+ 2\sigma_\varepsilon^2$}. This justifies the introduction of $\widehat{s}^2$.
	
	Note that $\hat r_{jj^\prime}^{\D}$ is still biased with respect to local noise (see Table~\ref{tab:summary}).
	
	A more formal proposition and proof for this estimator are provided in~\ref{sup:s:prop_proof_D}.


	\subsubsection{Replicates and use of a priori disconnected regions: method $\nRD$}
	\label{sec:RD}

	\jf{Combining replicates and the idea based on differences motivates us to propose the following estimator (see Sections~\ref{sec:R} and~\ref{sec:rD} for notation)
		\begin{equation}
			\widehat{r}^\RD_{jj^\prime}=\frac{1}{B} \sum_{b=1}^B \frac{\frac{1}{4}\sum_{\alpha,\beta=1}^2\widetilde{\cor}(\Vect{Y}_{i_\alpha^{(b)}},\Vect{Y}_{{i^\prime}_\beta^{(b)}};\Vect{Y}_{{k}^{(b)}},\Vect{Y}_{{k}^{\prime(b)}} )}{\sqrt{|\widetilde{\cor}(\Vect{Y}_{i_1^{(b)}},\Vect{Y}_{i_2^{(b)}} ; \Vect{Y}_{k^{(b)}},\Vect{Y}_{{k^\prime}^{(b)}}) \; \widetilde{\cor}(\Vect{Y}_{{i^\prime}_1^{(b)}},\Vect{Y}_{{i^\prime}_2^{(b)}} ; \Vect{Y}_{k^{(b)}},\Vect{Y}_{{k^\prime}^{(b)}})   |}  }
			\label{eq.srave}
		\end{equation}
		It is worth pointing out that $r^\RD_{jj^\prime}$ is independent of $\sigma_\varepsilon$ and $\sigma_e$ and equals the unknown $r_{jj^\prime}$ if $\rho_\delta$ is close to 1.}
		A more formal proposition and proof for this estimator are provided in~\ref{sup:s:prop_proof_RD}.

	
	\subsection{Localized versions of inter-correlation estimators}

	As mentioned previously, when noisy signals are averaged, the signal to noise ratio increases. A very popular method in neuroimaging analyses is to apply a Gaussian smoothing on the fMRI volumes~\citep{worsley_three-dimensional_1992,worsley_unified_1996,poline_combining_1997}. However, applying a large kernel width may have dramatic effect on brain connectivity~\citep{triana2020effects}. Some earlier work on PET connectivity used a local neighbourhood centered around voxels of interest to smooth the signal in each region prior to connectivity estimation~\citep{kohler_functional_1998}. We introduce in this section estimators using local neighborhoods to control the smoothing effect on correlation estimations.

	\subsubsection{Local correlation of averages  (method $\nlCA$)}\label{sec:lCA}
	
	\jf{Motivated by the first two estimators, we propose to estimate $r_{jj^\prime}$ using an empirical average of local spatial averages.} For $b=1,\ldots,B$, let $\mathcal{V}_j^{(b)}$ (resp. $\mathcal{V}_{j^\prime}^{(b)}$) be a $\nu$-neighborhood of $\r_j$ (resp. $\r_{j^\prime}$). \jf{We define}
	\begin{equation}
		\widehat{r}^{\lCA}_{jj^\prime}=\frac{1}{B} \sum_{b=1}^B\widehat{\cor}(\Vect{\bar{Y}}_{\mathcal{V}_j^{(b)}},\Vect{\bar{Y}}_{\mathcal{V}_{j^\prime}^{(b)}}).
		\label{eq.lave}
	\end{equation}

	\subsubsection{Local average of replicates (method $\nlR$)} \label{sec:rlR}
	
	This estimator consists in replacing single indices by neighborhoods in~\eqref{eq.rave}. For $b=1,\ldots,B$, let $\mathcal V_{j_1}\HL{^{(b)}}$ and $\mathcal V_{j_2}\HL{^{(b)}}$ (resp. $\mathcal V_{j_1^\prime}\HL{^{(b)}}$ and $\mathcal V_{j_2^\prime}\HL{^{(b)}}$) be two $\nu$-neighborhoods in $\r_j$ (resp. $\r_{j^\prime}$) such that 
	for any $i_1^{(b)} \in \mathcal V_{j_1}^{(b)}$, $i_2^{(b)} \in \mathcal V_{j_2}^{(b)}$, $| i_1^{(b)} - i_2^{(b)} | = \delta \geq p $ (resp.  $| i_1^{\prime (b)} - i_2^{\prime (b)} | = \delta \geq p$ for any $i_1^{\prime (b)} \in \mathcal V_{j_1^\prime}^{(b)}$, $i_2^{\prime (b)} \in \mathcal V_{j_2^\prime}^{(b)}$).
	The local average of replicates based estimator is defined by
		\begin{equation}
			\widehat{r}_{jj^\prime}^{\lR}=\frac{1}{B} \sum_{b=1}^B \frac{\frac{1}{4}\sum_{\alpha,\beta=1}^2\widehat{\cor}(\Vect{\bar Y}_{\mathcal{V}_{j_\alpha}^{(b)}},\Vect{\bar Y}_{\mathcal{V}_{j^\prime_\beta}^{(b)}})}{\sqrt{
					|
					\widehat{\cor}(\Vect{\bar Y}_{\mathcal{V}_{j_1}^{(b)}},\Vect{\bar Y}_{\mathcal{V}_{j_2}^{(b)}}) \; \widehat{\cor}(\Vect{\bar Y}_{\mathcal{V}_{j^\prime_1}^{(b)}},\Vect{\bar Y}_{\mathcal{V}_{j^\prime_2}^{(b)}})
					|
				}
			}.
			\label{eq.lrave}
		\end{equation}

	\subsubsection{Local averages and use of disconnected regions (method $\nlD$)}
	
	We use in particular notation introduced in Sections~\ref{sec:rlR} and~\ref{sec:rD} to propose the estimator $\widehat{r}^\lD_{jj^\prime}$ given by
	\begin{equation} \label{eq.slave}
		\jf{\widehat{r}^\lD_{jj^\prime}} = \frac{1}B \sum_{b=1}^B \widetilde{\cor} ( \Vect{\bar Y}_{\mathcal{V}_{j}^{(b)}}, \Vect{\bar Y}_{\mathcal{V}_{j^\prime}^{(b)}};\Vect{\bar Y}_{\mathcal{V}_{k}^{(b)}},\Vect{\bar Y}_{\mathcal{V}_{k^\prime}^{(b)}} ).
	\end{equation}

	\subsubsection{Replicates, local averages and use of a priori disconnected regions (method $\nlRD$)}
	
	This estimator is a local version of $\widehat{r}_{jj^\prime}^{\RD}$ and is defined by
		\begin{equation}
			\widehat{r}_{jj^\prime}^\lRD=\frac{1}{B} \sum_{b=1}^B \frac{\frac{1}{4}\sum_{\alpha,\beta=1}^2\widetilde{\cor}(\Vect{\bar Y}_{\mathcal{V}_{j_\alpha}^{(b)}},\Vect{\bar Y}_{\mathcal{V}_{{j^\prime_\beta}}^{(b)}}  ;\Vect{\bar Y}_{\mathcal{V}_{k}^{(b)}},\Vect{\bar Y}_{\mathcal{V}_{k^\prime}^{(b)}} )}{\sqrt{ 
					\widetilde{\cor}(\Vect{\bar Y}_{\mathcal{V}_{j_1}^{(b)}},\Vect{\bar Y}_{\mathcal{V}_{{j}_2}^{(b)}};\Vect{\bar Y}_{\mathcal{V}_{k}^{(b)}},\Vect{\bar Y}_{\mathcal{V}_{k^\prime}^{(b)}} )  \;
					\widetilde{\cor}(\Vect{\bar Y}_{\mathcal{V}_{j_1^\prime}^{(b)}},\Vect{\bar Y}_{\mathcal{V}_{{j^\prime}_2}^{(b)}};\Vect{\bar Y}_{\mathcal{V}_{k}^{(b)}},\Vect{\bar Y}_{\mathcal{V}_{k^\prime}^{(b)}} )
			}  }.
			\label{eq.slrave}
		\end{equation}
	
		\newcommand{\plus}{{\Large $+$}}
	\newcommand{\minus}{{\Large $-$}}
	\newcommand{\plusminus}{{\Large $\pm$}}

	\subsection{Summary of estimators}

	We have formalised 9 estimators for inter-region correlation in fMRI, 6 of which are novel to the best of our knowledge. They vary in terms of their theoretical sensitivity to three factors:  differences in region sizes and region intra-correlations ($\bar \rho_{\r_j}\ll 1$), local noise ($\sigma_\varepsilon$), and global noise ($\sigma_e$). Table~\ref{tab:summary} summarises estimators properties qualitatively using \minus \; for estimators that are sensitive to these factors, \plus \; for estimators that are insensitive, and \plusminus \; for those that are in-between. The \nCA, \nAC, \nR, and \nD \; estimators are also sketched in Figure~\ref{fig:illustrationEstimators}. 
	
	As an example, let us interpret the properties of \nCA \; shown in Table~\ref{tab:summary} in terms of these factors. First, we observe that the limit $\widehat{r}^\CA_{jj^\prime}/r^{\CA}_{jj^\prime}$ strongly depends on the region size. Indeed, even in absence of noise this limit is $1/\sqrt{\bar \rho_{\r_j}\bar \rho_{\r_{j^\prime}}}$, which can be quite far from 1 especially for very large regions (so the estimator is sensitive to local noise and denoted  \minus \; in the corresponding column).
	Now imagine that  $\bar \rho_{\r_j}\bar \rho_{\r_{j^\prime}}=1$ and that $\sigma_e^2=0$ then the limit becomes 
	$1/
	\sqrt{
	(1+\sigma_{\varepsilon,j}^2 \bar\eta_{\r_j})
	(1+\sigma_{\varepsilon,j^\prime}^2 \bar\eta_{\r_j^\prime} )}  
	$. Since it is expected that $\bar \eta_E$ is small (see \eqref{eq:epsbarE}), especially for large sets $E$, this limit should be quite close to 1 in this situation (\plus).
	Finally, if $\bar \rho_{\r_j}\bar \rho_{\r_{j^\prime}}=1$ and $\sigma_\varepsilon^2=0$, and assume for simplicity that $\sigma_j=\sigma_{j^\prime}=1$, then $\widehat r^{\CA}_{jj^\prime}$ would converge towards $(r_{jj^\prime} + \sigma_e^2)/(1+\sigma_e^2)$ which can significanlty deviate from $r_{jj^\prime}$ when the global noise is strong (\minus).

	This does not describe at all finite sample properties of the different estimators. Obviously, we could be tempted to always use the last two estimators (methods $\RD$ ad $\lRD$) which seem to be the most robust to additional noises. However, it is clear that the price to pay for reaching \HL{robustness} is accuracy. We propose to \HL{investigate} these finite sample properties in a simulation study (Section~\ref{sec:DataSim}) and real datasets (Sections~\ref{sec:DataRat} and~\ref{sec:DataHCP} ).
	
	We note that evaluating asymptotic variances of the different estimators would add too much notation, assumptions and \HL{technicalities}, and is left for future work.
	
	R code implementing all estimators, as well as NIFTI time-series extraction and preprocessing, is available at \url{https://gitlab.inria.fr/q-func/ireco4fmri}

\begin{landscape}
		\begin{table}
			\centering
			\begin{tabular}{|p{6cm}|c|p{2cm}p{1cm}p{1cm}|}
				\hline
				Estimator &\multicolumn{1}{|p{5cm}|}{Limit \quad $r^\bullet_{jj^\prime}/ r_{jj^\prime}$}& \multicolumn{3}{c|}{Sensitivity to }\\
				 &  & 
				$\bar \rho_{\r_j}\ll 1$
				&$\sigma_\varepsilon$  &$\sigma_e$ \\
				\hline \rule{0pt}{20pt} 
				$\widehat{r}^{\CA}$ (see \eqref{eq.agg})&
				$
				\frac{1 + \sigma_{e,jj^\prime}^2/r_{jj^\prime} }
				{
					\sqrt{
						( \bar \rho_{\r_j} +   \sigma_{\varepsilon,j}^2 \bar \eta_{\r_j}+\sigma_{e,j}^2) 
						( \bar \rho_{\r_{j^\prime}} +  \sigma_{\varepsilon,j^\prime}^2 \bar \eta_{\r_{j^\prime}} +\sigma_{e,j^\prime}^2)
					}
				}$
				&\minus&\plus & \minus \\ 
				\hline \rule{0pt}{20pt}
				$\widehat{r}^{\AC}$ (see \eqref{eq.ave}) & 
				$\frac{1+ \sigma_{e,jj^\prime}^2/r_{jj^\prime} }{
					\sqrt{	(1  + \sigma_{\varepsilon,j}^2  +\sigma_{e,j}^2)
						(1 + \sigma_{\varepsilon,j^\prime}^2 +\sigma_{e,j^\prime}^2) }
				}$ & 
				\plus & \minus & \minus \\
				\hline \rule{0pt}{20pt}
				$\widehat{r}^{\lCA}$ (see \eqref{eq.lave})&
				$\frac{1 + \sigma_{e,jj^\prime}^2/r_{jj^\prime} }
				{
					\sqrt{( \bar \rho_{\mathcal V} +   \sigma_{\varepsilon,j}^2 \bar \eta_{\mathcal V}+\sigma_{e,j}^2)
						( \bar \rho_{{\mathcal V}} + \sigma_{\varepsilon,j^\prime}^2 \bar \eta_{{\mathcal V}} +\sigma_{e,j^\prime}^2)}
				}$& \plus & \plusminus&\minus\\
				\hline \rule{0pt}{20pt}
				$\widehat{r}^{\R}$ (see \eqref{eq.rave})&
				$\frac{1 + \sigma_{e,jj^\prime}^2/r_{jj^\prime}}{
					\sqrt{ | (\rho_\delta + \sigma_{e,j}^2)
						(\rho_\delta + \sigma_{e,j^\prime}^2) | } 
				}$
				&\plus&\plus&\minus\\
				\hline \rule{0pt}{20pt}
				$\widehat{r}^{\lR}$ (see \eqref{eq.lrave})&
				$\frac{1 + \sigma_{e,jj^\prime}^2/r_{jj^\prime}}{
					\sqrt{ | (\bar \rho_{\mathcal V, \mathcal V^\prime,\delta} + \sigma_{e,j}^2)
						(\bar\rho_{\mathcal V, \mathcal V^\prime,\delta} + \sigma_{e,j^\prime}^2) | }} $
				&\plus&\plus&\minus\\
				\hline \rule{0pt}{20pt}
				$\widehat{r}^{\D}$ (see \eqref{eq.save})&
				$\frac{1 }{
					\sqrt{	\left( 1 +  \sigma_{\varepsilon,j}^2 \right) \left( 1 +  \sigma_{\varepsilon,j^\prime}^2 \right) }
				}$
				&\plus&\minus&\plus\\
				\hline \rule{0pt}{20pt}
				$\widehat{r}^{\lD}$ (see \eqref{eq.slave})&
				$\frac{1 }{
					\sqrt{(\bar \rho_{\mathcal V} +\sigma_{\varepsilon,j}^2 \bar \eta_{\mathcal V})
						(\bar \rho_{\mathcal V} +\sigma_{\varepsilon,j^\prime}^2 \bar \eta_{\mathcal V})}
				}$
				&\plus&\plusminus&\plus\\
				\hline \rule{0pt}{20pt}
				$\widehat{r}^{\RD}$ (see \eqref{eq.srave})&
				$\frac {1}{|\rho_{\delta}|}$
				&\plus&\plus&\plus\\
				\hline \rule{0pt}{20pt}
				$\widehat{r}^{\lRD}$ (see \eqref{eq.slrave})&
				$\frac{1}{|\bar\rho_{\mathcal V, \mathcal V^\prime,\delta}|}$
				&\plus&\plus&\plus\\
				\hline 
			\end{tabular}
			\caption{Expected limits and properties for existing and novel estimators of inter-correlation $r_{jj^\prime}$, under the model~\eqref{eq:model}. We refer the reader to Section~\ref{sec:estimators} for details on notation. In particular $\sigma_{e,j}^2$, $\sigma_{\varepsilon,j}^2$ and $\sigma_{e,jj^\prime}^2$ are given by \eqref{eq:morenotation} while $\bar \rho_E, \bar \eta_E $ and $\bar \rho_{E,E^\prime}$ for two sets of indices $E,E^\prime$ are given by~\eqref{eq:defrhoetaagg} and~\eqref{eq:rhoEEprime}. Sensitivity of estimators to three factors are reported: differences in region sizes and region intra-correlations ($\bar \rho_{\r_j}\ll 1$), local noise ($\sigma_\varepsilon$), and global noise ($\sigma_e$). Estimators that are sensitive to these factors are denoted \minus \; those that are insensitive are denoted \plus \; and those in-between are denoted insensitive \plusminus \; .}
			\label{tab:summary}
		\end{table}
	\end{landscape}

\subsection{Datasets}
We used three different datasets to evaluate our estimators: a simulated dataset, a rat dataset including both dead and live animals, and a healthy human subject dataset with test-retest data.

\subsubsection{Simulated data}
\label{sec:DataSim}

For different sets of parameters (local noise, global noise), we generated according to model~\eqref{eq:model} two regions containing 20 and 40 voxels, respectively, each corresponding to a time series of length 1000,
resulting in a total of 60 time series. This simulation was therefore done in dimension 1 to save time and we abusively keep the terminology voxel for a one-dimensional cell.
We assumed the signal, local noise and global noise followed Gaussian distributions.
 Throughout the simulations the true inter-correlation was set to $0.6$ and the local noise $\varepsilon_i(t)$ and $\varepsilon_{i^\prime}(t)$ were assumed to be uncorrelated. 
 The intra-correlation was defined by the following spatial model: $\rho_{|i^\prime-i|} = \max( 1- |i^\prime-i|/100, 0.6)$. This corresponds to a setting where two voxels inside a given region have moderate intra-correlation.

\subsubsection{Rats data}
\label{sec:DataRat}

Using a 9.4T machine (Paravision 6.0.1, Bruker, Ettlingen, Germany), fMRI data were acquired for both dead and alive rats in \cite{Pawela2008}.
Twenty-five rats were scanned and identified in 4 different groups: DEAD, ETO-L (Etomidate), ISO-W (Isoflurane) and MED-L(Medetomidine).
The first group contains dead rats and the three last groups correspond to different anesthetics. In this paper, we show results with data from three rats, one dead and two alive with different anesthetics (ETO-L, ISO-W).

The duration of the scanning was 30 minutes, using single-shot echo-planar imaging with TR / TE =  500 / 20 ms, so that $3600$ time points were available at the end of experiment. The resolution was $0.47  \times 0.47 \times 1.00$ mm, slice gap 0.1 mm, 9 slices. After preprocessing as explained in \cite{guillaume2020functional}, $51$ brain regions for each rat were extracted using an in-house atlas. Sufficiently large regions are needed to be able to use the $\nR$ estimator. We hence discarded regions that contained fewer than 40 voxels, and were left with $18$ brain regions: The anterior cingulate cortex (ACC), bilateral Insular cortex (Ins\_r and Ins\_l), bilateral primary motor cortex (M1\_r and M1\_l), bilateral somatosensory 1 (S1\_r and S1\_l), bilateral somatosensory 1 barrel field (S1BF\_r and S1BF\_l), bilateral auditory cortex (AU\_r and AU\_l), bilateral caudate-putamen (striatum) (CPu\_r and CPu\_l), bilateral thalamus (Th\_r and Th\_l), bilateral basal forebrain region (BF\_r and BF\_l), bilateral hippocampus (HIP\_r and HIP\_l).

Voxel time series were waveled-filtered using Daubechies orthonormal
          compactly supported wavelet of length 8. 


The pre-extracted, wavelet-filtered time series for the rat data are available at \url{https://dx.doi.org/10.5281/zenodo.7254133}.

\subsubsection{Human Connectome Project data}
\label{sec:DataHCP}

We also evaluated our estimators on a subset of the Human Connectome Project (HCP) Young Adult 1200 Subjects release, WU-Minn Consortium pre-processed~\citep{glasser_minimal_2013} (connectome db data package \textit{Resting State fMRI 1/2 Preprocessed}). We selected 36 subjects with two rs-fMRI acquisitions on different days. The TR was 720 ms and the duration of acquisition was 14 min and 24s.

The preprocessed fMRI data was segmented into 89 regions with SPM \textit{New Segment} using a modified AAL template: merging some of the regions, reducing the parcellation
to 89 regions. Merged regions are: frontal medial orbital and rectus (one region
for left and one for right hemisphere); occipital superior, middle and inferior (one region for left and one for right hemisphere); temporal pole superior and medial (one region for left and one for right hemisphere); the cerebral crus (one region for left and one for right hemisphere); areas III, IV, V and VI of cerebellum (one region for left and one for right hemisphere); areas VII, VIII, IX, X of cerebellum (one region for left and one for right hemisphere) and finally, the
vermis (one single region for both hemispheres). Other details are available in~\cite{termenon2016reliability}.

Voxel time series were wavelet filtered using Daubechies orthonormal
          compactly supported wavelet of length 8.

The code to extract the time series from the preprocessed HCP data is available at \url{https://gitlab.inria.fr/q-func/ireco4fmri}.

\subsection{Evaluation and metrics}

Preferable inter-correlation estimators have three properties i) face validity, ii) good repeatability, iii) preservation of the differences between individuals (discriminative power), and iv) independence from region size.

First, on \textit{simulated data}, we qualitatively inspected the bias and variance of the distribution of correlation values with respect to known ground truth  for various levels of global and local noise.

Then, using \textit{rat data}, we performed a face validity analysis of the estimators, with the premise that dead rats should show no functional connectivity (the correlation distribution should be centered at zero). In order to quantify the differences between correlation
values obtained for dead and live rats, we computed the Wasserstein distance between the correlation distributions of each anesthetized rat in comparison to that of a dead rat. A low value of the Wasserstein distance indicates that correlations values of live and dead rats are comparable and counts negatively in the evaluation of an estimator. 

To evaluate the repeatability of the proposed estimators on the rat dataset, we split the time series in two equal parts. We computed the correlations on each part using the whole range of proposed estimators, and computed the Concordance Correlation Coefficient (CCC)~\citep{lin_concordance_1989} between splits to provide a scaled measure of agreement, where 1 is perfect agreement and 0 is no agreement. A preferable estimator should be more repeatable and have higher CCC.

To quantify the similarity of connectivity graphs between estimators, we computed the number of common edges between graphs obtained from each estimator. To this end we used a sparsity threshold equal to 20\% of the total number of edges (i.e., 27  edges in our case with 18 regions).

For \textit{human data}, we used rs-fMRI sessions from different days, in order to evaluate reproducibility of correlation coefficients. This was analysed using CCC, again with a preferable estimator being more reproducible and having higher CCC. 

We also evaluated the reproducibility of graph metrics between sessions.  To this end we used a sparsity threshold equal to 20\% of the total number of edges, keeping only edges with the highest correlation (i.e., 783 edges in our case with 89 regions), and binarized the edges. In order to compute graph metrics, we forced the graph to be connected by applying a minimum spanning tree~\citep{alexander-bloch_disrupted_2010}. Then we computed classical graph metrics: betweenness centrality, transitivity, global and local efficencies using package iGraph. Reproducibility was evaluated using the CCC.

We also summarized the differences of connectivity graphs between estimators, by computing the number of common edges between graphs obtained from $\nCA$ and $\nlCA$ using thresholding at the 20th percentile  (i.e., 783 edges with 89 regions), and visualized the difference qualitatively by taking absolute values of correlation values for each estimator, rank-transforming, and computing median difference in ranks across all subjects, 

Additionally, we also evaluated discriminative power of the various estimators via three metrics: inter vs intra-subject graph distance, a non-parametric test of the same, and identification rate using functional connectome fingerprinting~\citep{finn_functional_2015}. A desirable estimator should provide estimates that preserve inter-individual differences. 

We defined the intra-subject distance as the distance beween the graph representing the first rs-fMRI session and the graph representing the second rs-fMRI session. The inter-subject distance was computed between each subject's first session and all other subject's first sessions. Separation between the intra-subject distances and the inter-subject distances was quantified by mean and standard deviation of the distributions, and by a Wilcoxon rank-sum test on multiple random splits of subject data, to avoid having multiple measurements of the same subjects. Here, we repeated 10 times the following procedure for each estimator of interest: first, split the subjects into two disjoint sets - one used to compute intra-distances (18 subjects), and one to compute inter-distances (18 subjects). Within the inter-distances set, 9 subject pairs were formed randomly. Then, a one-sided Wilcoxon rank-sum hypothesis test for the null hypothesis of no difference between inter- and intra-distances was performed, yielding a W statistic and a p-value for each of the 10 runs. We then computed the average W value across runs, as well as the harmonic mean p-value~\citep{wilson_harmonic_2019} across runs, a procedure with strong family-wise error rate (FWER) control even for positively dependent tests.

To compute identification rate, functional connectome fingerprinting represent each subject's graph $g$  as a vectorized version $\mathbf{a}$ of the upper-triangular (or lower-triangular) part of the full inter-region correlation matrix (whose entries are $r_{ii^\prime}$), and computes the fingerprinting distance between graphs as $d(g_1,g_2)=1- \widehat{\cor}(\mathbf{a}_1,\mathbf{a}_2)$, where 
$\cor$ denotes Pearson correlation. From the (intra, inter) fingerprinting distance distributions, the identification counts as correct if the intra-subject distance is lower than all inter-subject distances. This is equivalent to a top-1 recognition rate. We note there are many other possibilities to compute distances between such brain graphs~\citep{richiardi_machine_2013,ng_transport_2016,dadi_benchmarking_2019}, including computing distances between graph embeddings, which could substantially alter results.

Finally, we evaluated the dependence on region size by computing Spearman correlations between atlas region size and the average of correlations in which the region is involved (itself averaged across subjects). A preferable estimator should minimize dependence to region size, and show lower Spearman correlation. We tested differences between estimators using a paired t-test between these Spearman correlations. 

\section{Results}


	\subsection{Evaluation on simulated data} \label{sec:simul}

\begin{figure}[H]
		\centering  
		\includegraphics[width=\linewidth]{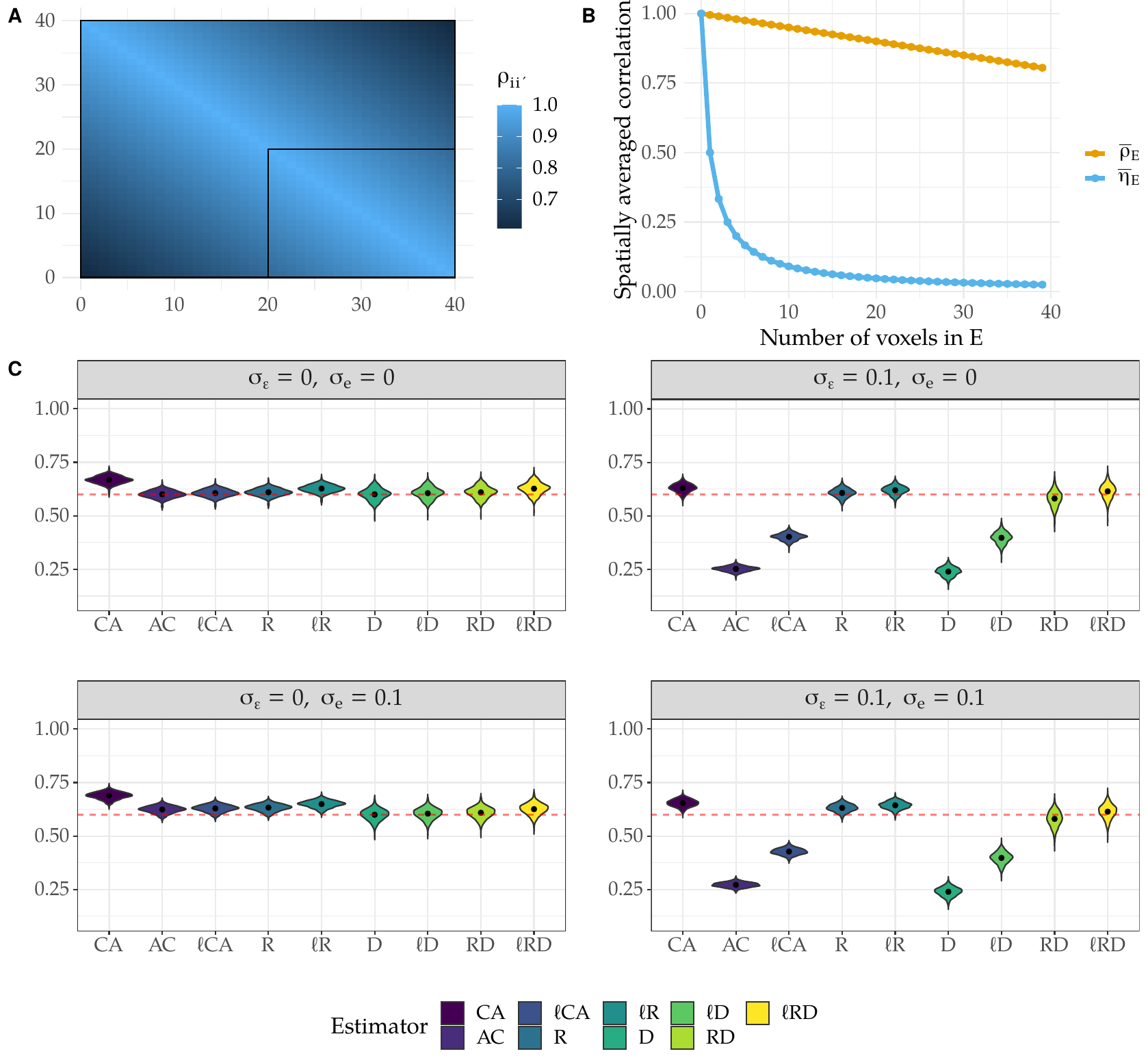}
		\caption{Simulation setup and results. (A) The two simulated one-dimensional regions (one with 40 ``voxels'', the other with 20 ``voxels'', shown as an inset) and their intra-correlation structure. Intra-correlation $\rho_{ii^\prime}$ decays with distance. (B) Intra-correlation (vertical axis) as a function of the size of region $E$ (horizontal axis). $\bar{\rho}_E$ (orange): average intra-correlation of signal, $\bar{\eta}_E$ (blue): average intra-correlation of noise. Average noise intra-correlation decays sharply with region size. (C) Estimates of the inter-correlation parameter $r$ between two regions, based on 500 simulation runs of the general model~\eqref{eq:model}. Situations with no noise, local noise or globa\HL{l} noise are considered. The true inter-correlation is depicted by the red dashed line.}
		\label{fig:boxplots}
	\end{figure}

	We first study the properties of the estimators in the absence of noise ($\sigma_\varepsilon=\sigma_e=0$). Figure~\ref{fig:boxplots} shows boxplots based on $500$ replications of the general model~\eqref{eq:model} with parameters described previously. Figure~\ref{fig:boxplots} (with $\sigma_\varepsilon=\sigma_e=0$) shows that when the intra-correlation within each region $\r_j$ and $\r_{j^\prime}$ is high for any pair of voxels, then the estimates of $r$ are quite satisfactory even if we can already observe  a slight bias for the method  $\nCA$. 
	It is to be noticed that in terms of variance the methods based on replicates (i.e. methods $\nR$, $\nlR$) are as efficient as the methods $\nAC$ and $\nlCA$. The methods based on the a priori knowledge of two other disconnected regions (methods based on ``differences'', i.e. the methods $\nD$, $\nlD$, $\nRD$ and $\nlRD$) have higher dispersion than the other ones. Finally, we also observe that the ``complex'' methods $\nlR$ and $\nlRD$ based on local averages generate a more important bias than other methods. This bias is clearly smaller than the one observed for the method $\nCA$ and illustrates the following fact: when we suspect that the intra-correlation matrix is not very well concentrated on the diagonal then the size of the neighborhood ($\nu$ in the definition of the estimators) should be chosen sufficiently small.

	We now turn to study the influence of local and global noise\HL{, which} are also illustrated in Figure~\ref{fig:boxplots}. The variances of local 
	and global noise are parameterized by fixing the signal-to-noise ratio: when for instance $\sigma_e=0$, we fixed $\sigma_\varepsilon^2$ as follows  
	\[
	\SNR_\varepsilon = 10 \log_{10}  \left( \frac{\min(\sigma_j^2,\sigma_{j^\prime}^2)}{\sigma_\varepsilon^2}\right)
	\Leftrightarrow
	\sigma_\varepsilon^2 =  10^{-\SNR_\varepsilon/10} \min(\sigma_j^2,\sigma_{j^\prime}^2).
	\]
	When $\sigma_\varepsilon\neq 0$ and $\sigma_e=0$, as expected, Figure~\ref{fig:boxplots} shows that  the methods based on replicates are able to estimate correctly the inter-correlation parameter $r$ and these methods remain efficient whatever the value of the $\SNR$. 
	The methods $\nAC$, $\nlCA$, $\nD$ and $\nlD$ are strongly affected by this additional local noise and exhibit a high negative bias. As explained in Section~\ref{sec:CA}, the method $\nCA$ which averages the signals in the regions $\r_j$ and $\r_{j^\prime}$ is able to reduce the effect of the local noise.


	\subsection{Evaluation on rat data}

\begin{figure}
	    \centering
	    \includegraphics[width=\textwidth]{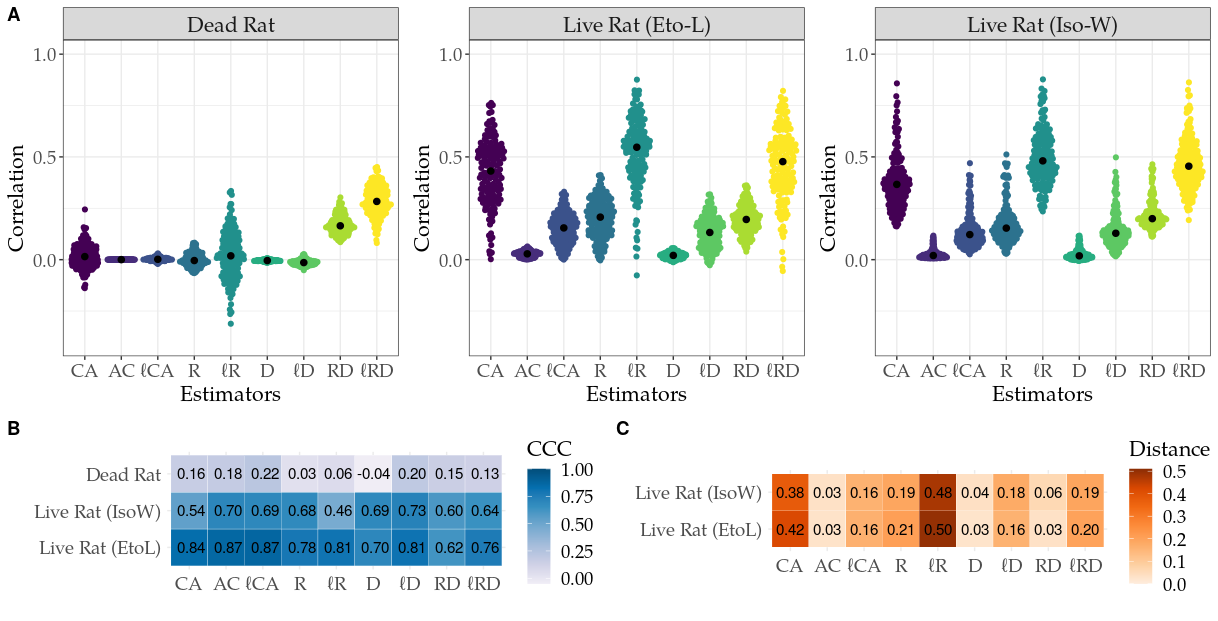}
	    \caption{Rat data results. \textbf{A.} Empirical distribution of the correlation estimators for all pairs of brain regions for a dead and two anesthetized rats, for all proposed estimators. In the dead rat, the correlation of averages (CA) estimator is providing high values where null correlations should be observed. For the live rat the average of correlation estimator (AC) is providing very low values where non null correlations should be observed. \textbf{B.} The Concordance Correlation Coefficient (CCC) 
	    for the repeatability of the different estimators for all rats, calculated between the first and second half of the BOLD time series.
		Higher CCC 
		corresponds to a more repeatable estimator.
		\textbf{C.} Wasserstein distances between the correlation distribution of each anesthetized rat and that of the dead rat, for all estimators. $\AC$, $\D$, $\RD$ 
		have a very low distance, indicating that correlation values are similar between dead and live rats for these estimators.}
	    \label{fig:rats}
	\end{figure}

        Figure~\ref{fig:rats}(A) shows the correlation values obtained on rats for all pairs of brain regions, 153 
        in our case. For this data set, we know that for the dead rat we are under the full null hypothesis as no legitimate functional activity should be detected. Thus the estimated correlations should be close to zero. This is the case for estimators $\nAC$, $\nR$, $\nlCA$, $\nD$ and $\nlD$. However, the other estimators showcase a clear bias towards positive values. The method $\nCA$ namely yields unexpectedly high values of correlations. These correlations correspond to regions that are close together~\citep{becq_10.1088/1741-2552/ab9fec}. In order to validate these methods, we also apply our estimators to live rats. The results of two live rats is shown in Figure~\ref{fig:rats} (A, right). As expected, due to the local noise, the methods $\nAC$ and $\nD$ do not provide satisfactory results as the correlation values are very close to zero. One of the best method in this case is $\nlCA$, where sufficient non-zero correlations are obtained. Wasserstein distance computations (Figure~\ref{fig:rats} (C)) show that $\nAC$, $\nD$, and $\nRD$ have the lowest Wasserstein distance values, indicating that the correlation distribution of the live rats resemble that of a dead rat.
	
        Figure \ref{fig:rats}(B) shows Concordance Correlation Coefficient results. Consistent with the all-noise nature of the data, the dead rat exhibited very low repeatability, with $\nlCA$ providing the highest at 0.22. On the live Eto-L rat, estimators had approximately the same repeatability, with $\nRD$ showing the lowest CCC at 0.62 and $\nAC$ tied with $\nlCA$ for highest at 0.87. For the Iso-W rat, $\nlR$ had the lowest CCC at 0.46, $\nCA$ the second lowest at 0.54, and $\nlD$ the highest at 0.73.
        
        Combining all of these results, $\nlCA$, $\nR$ and $\nlD$ hence seem to be the most adequate correlation estimators. However, as shown in formula \eqref{eq:lD}, the estimator $\nlD$ is difficult to implement. Indeed, it requires the definition of two other regions uncorrelated with the main brain regions of the parcellation and uncorrelated with themselves. Moreover, $\nR$ cannot be estimated when regions are too small, which is often the case in rat data. From now on, we will hence focus on estimator $\nlCA$. 

        We then quantified the edges in common between the networks obtained via the two estimators $\nCA$ (which is currently the most widely used estimator)  and $\nlCA$. For the dead rat, 67\%
        of edges are in common between the two estimators. Additionally, 60\% and 77\% of edges are similar for the live rats.

\subsection{Evaluation on human data}

Based on our findings on the rats datasets, we evaluate the performances of the three estimators $\nCA$ (most common estimator, highest dead-live rat distance), $\nAC$ (low dead-live rat distance) and $\nlCA$ (high dead-live rat distance) for 36 subjects of the HCP dataset.

\begin{figure}[htbp]
		\centering
		\includegraphics[width=\textwidth]{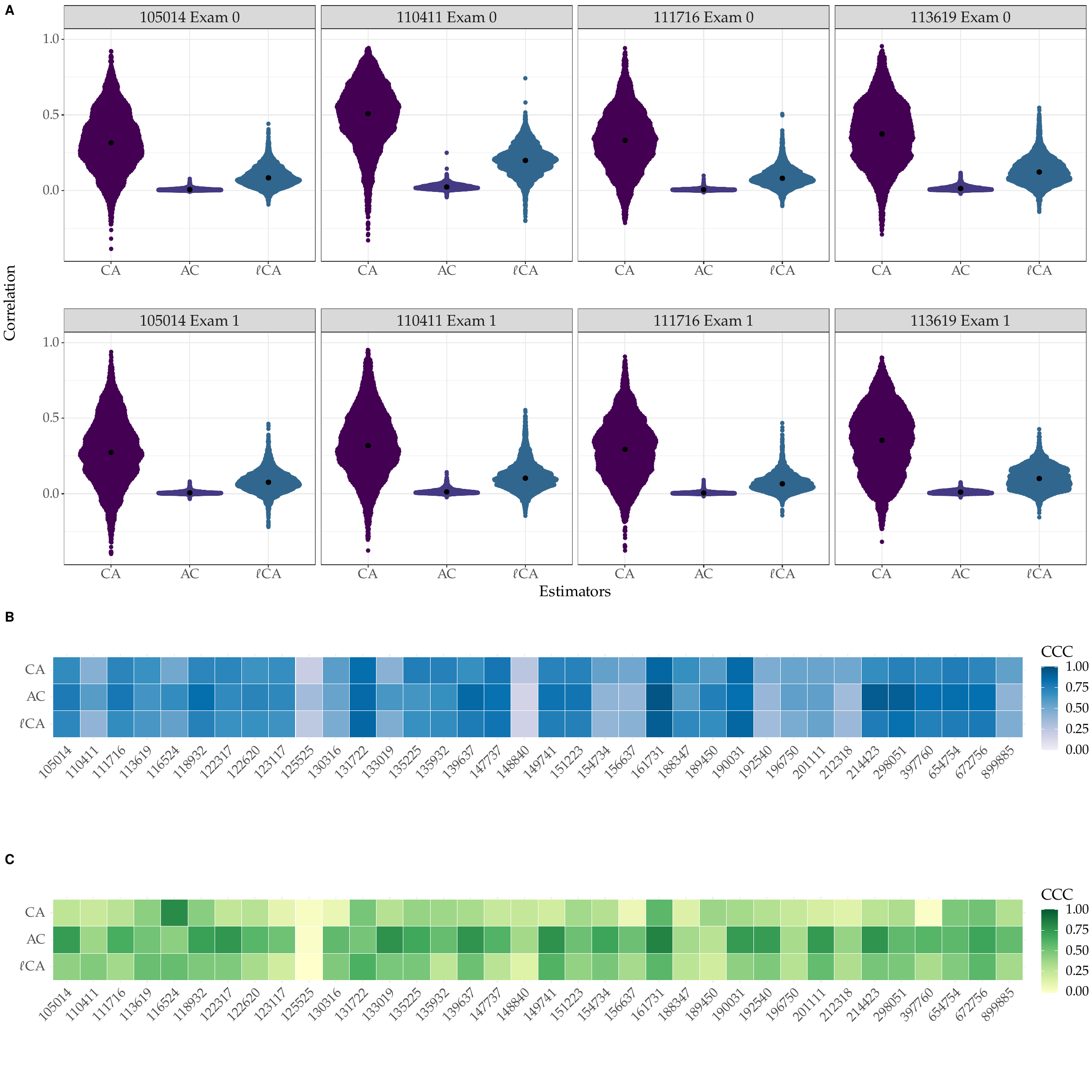}
		\caption{Human data results. \HL{\textbf{A.} Empirical distribution of inter-regional correlations for three selected estimators for all pairs of brain regions for four human subjects. Each subject was scanned twice, on different days. 
		\textbf{B.} Concordance correlation Coefficient (CCC) for the  reproducibility of the inter-regional correlation values obtained by different estimators for all human subjects, computed between the two examinations. Higher CCC indicates a more repeatable estimator. All estimators have broadly similar reproducibility. \textbf{C.} reproducibility of a topological graph metric (betweeness). Again all estimators give broadly similar results, with slightly higher reproducibility for $\nAC$}}
		\label{fig:HCP}
	\end{figure}

Figure~\ref{fig:HCP}(A) reports the correlation values among all pairs of regions for four randomly selected HCP subjects. Consistent with the rat results, the estimator $\nCA$ yields the largest values of correlations, estimator $\nAC$ yields very low values, while $\nlCA$ values are different from zero, but smaller that $\nCA$ values.

 Reproducibility results for correlation estimates are shown in Figure~\ref{fig:HCP} (B). The Concordance Correlation Coefficient was similar between estimators (average (sd) across 36 subjects for $\nCA$: 0.64 (0.15), $\nAC$: 0.67 (0.20), $\nlCA$: 0.63 (0.19)), with variations in reproducibility reflecting inter-subject variability more than differences between estimators. For graph metrics reproducibility, we report only the results with betweenness in Figure~\ref{fig:HCP} (C), since similar results are obtained with other metrics. Here, the methods differed more, with average (sd) across 36 subjects for $\nCA$: 0.30 (0.17), $\nAC$: 0.58 (0.18), $\nlCA$: 0.41 (0.15). $\nCA$ had significantly lower CCC than $\nlCA$ (T=-4.6, $p=6.1e^{-5}$).
 
\begin{figure}[htbp]
		\centering
		\includegraphics[width=0.8\textwidth]{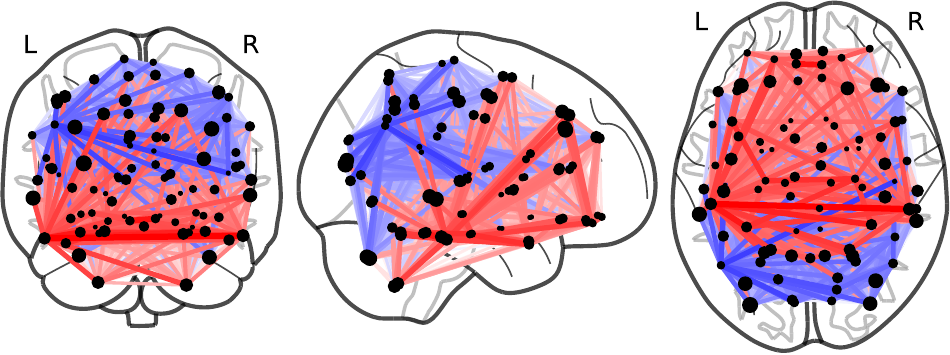}
		\caption{Largest differences between the $\nCA$ and $\nlCA$ estimators, median over 36 HCP subjects. Only the top 20\% differences are shown. Inter-regional correlations are taken in absolute value and  rank-transformed prior to computing differences (rank 1 for the strongest correlation, rank 2 for the second-strongest, and so on). Red indicates absolute correlations that are higher for the $\nlCA$ than the $\nCA$ estimator, while blue indicates the reverse. Node size is proportional to region size in the atlas. Estimator $\nCA$ on average shows hyperconnectivity in occipital and generally dorsal posterior regions, and hypoconnectivity in frontal, temporal, and general ventral anterior regions.}
		\label{fig:braingraph_diffs}
	\end{figure}

In the thresholded graphs, the percentage of edges in common between estimators $\nCA$ and $\nlCA$ was on average equal to 70\% for the thirty-six subjects used in this analysis for both sessions. Figure~\ref{fig:braingraph_diffs} shows median differences between the estimators in brain space across the HCP subjects.
 
 Looking at dependence on region size, the $\nCA$ estimator showed significantly more correlation with region size than the $\nlCA$ estimator (average (sd) across 36 subjects 0.54 (0.12) vs 0.39 (0.11), T=13.5, $p=1.8 e^{-15}$.

In terms of discriminative power between subjects, for connectome fingerprinting, $\nCA$ and $\nlCA$ achieved the same performance (72\% correct identification), while $\nAC$ had slightly lower performance (67\% correct identification). Group differences were also similar between estimators. Table~\ref{tab:fingerprinting} provides details.

\begin{table}[htbp]
\centering
\begin{tabular}{@{}ccccc@{}}
\toprule
estimator  & intra (sd)            & inter (sd) & W (p-value) & identification rate \\ \midrule
   $\nCA$  &  0.30 (0.12)      &  0.51 (0.14)  & -3.58 ($p_{hmp}=2.0e^{-4}$)     &   72\%                  \\
   $\nAC$  &  0.19 (0.12)      &  0.34 (0.14)  & -3.01 ($p_{hmp}=4.2e^{-4}$)    &    67\%                 \\ 
   $\nlCA$ &  0.27 (0.13)      &  0.46 (0.14)  & -3.31 ($p_{hmp}=2.1e^{-4}$)    &  72\%                   \\

   \bottomrule
\end{tabular}
\caption{Discriminative power of estimators on the human dataset. intra: Within-subject average and standard deviation of graph distances between first and second imaging session across 36 subjects; inter: same for between-subjects, using only the first session. W: average one-sided Wilcoxon rank-sum test value on 10 random splits, with corresponding harmonic mean p-value}
\label{tab:fingerprinting}
\end{table}

\section{Discussion}

In this paper we illustrate the effect of averaged data on estimators of correlation when two types of noises are present, local and global noise. The use of the classical correlation of averages is hindered by the presence of these noises in addition to the presence of
intra-correlations.
We proposed alternative estimators including correction terms to compensate the intra-correlations, local and global noises.
The performance of these estimators was evaluated on simulations, rats data, and human data, yielding several observations.
        
\subsection{The correlation of averages estimator is highly biased}
        
        The CA estimator tends to be highly biased, as illustrated on synthetic data where the ground truth is known, but also compared to other estimators, as shown on live rats and human data, where the mode of the distribution of correlation values is systematically among the highest found. We hypothesise that this is driven by a combination of low intra-correlation and large region sizes, which further lowers intra-correlation. This can be seen from the estimator definition in Eq.~\ref{eq.agg}. We also note that the $\nlCA$ estimator effectively reduces this influence of region size.
        
        In addition, Figure~\ref{fig:braingraph_diffs} revealed a systematic spatial bias between the $\nCA$ and $\nlCA$ estimator, exhibiting dorsal posterior hyper-connectivity for $\nCA$, and corresponding ventral anterior hypo-connectivity. The figure also suggests that the largest differences between the two estimators comes between regions that are the largest, further highlighting the reduced dependency to region size for the $\nlCA$ estimator. The spatial distribution of these differences suggests that caution is in order when examining large-scale resting-state networks derived from the $\nCA$ estimator, as some apparent topological properties of brain networks, such as modularity, could be driven in part by region size and region intra-correlation. Indeed, in our experiments, thresholded graphs differed in a large proportion of edges, both in rats (around 30\%-50\% edge differences) and humans (around 30\% edge differences).

\subsection{local noise and intra-correlation link to long-range correlation}

        In this paper, we explain the bias observed in $\nCA$ estimator by introducing hypotheses on both intra-correlation and noise. Indeed, previous studies on regional homogeneity~\citep{zang_regional_2004} showed relevant results on classification of pathologies based only on intra-regional properties. This was confirmed by a recent work on classification of intra-correlation~\citep{petersen_quantifying_2016} using Wasserstein distances. Based on these findings, we hypothesize that bias observed on inter-correlation is driven by intra-correlation and noise. Our simple simulation model illustrates the effect of local noise and intra-correlation. This is clearly displayed in Figure~\ref{fig:boxplots}, where the boxplots for the various estimators are plotted. However, it is important to note that under local noise, in this framework with controlled intra-correlation, estimator $\nCA$ is relatively close to the exact value. This may be explained by a trade-off in the denominator of the limit as expressed in Table~\ref{tab:summary}. In our simulation, we also observed that the $\nCA$ estimators bias depends on the intra-correlation and local noise. Indeed, high values of $\nCA$ tends to be observed when low values of intra-correlation are observed. These low values of intra-correlation have already been mentioned in the study of dynamics of neural networks~\citep{deco_how_2014} where local decorrelation was reported in real datasets. In our paper, for the first time, we proved a statistical explanation of the link between local decorrelations and long-range correlations using aggregated time series.
        
        The model chosen in this paper for intra-correlation and local noise was driven by statistical motivations to be able to write explicit formulas for the limit of the estimators. However, as observed in~\cite{jiang_regional_2016, deco_how_2014}, these hypotheses are realistic for resting-state fMRI data, where local decorrelations are observed. These local decorrelations can come from two factors: a low intra-correlation as modeled by the choice of the intra-correlation coefficients of the matrix, or a strong local noise. The stationary assumptions may be not adequate based on real data observations, however, this is compulsory to get simple interpretations of the limits of the estimators.

\subsection{Repeatability and reproducibility}

Repeatability of correlation values in dead rats was very low for all estimators, consistent with the random nature of the data. For live rats, the CCC ranged from 0.46 to 0.87 depending on specimen and estimator. For humans, 
 $\nCA$ and $\nlCA$ showed approximately the same reproducibility ($0.63$ average ($0.2$)), and $\nAC$ was slightly superior ($0.67$ average ($0.2$)). But reproducibility differences between estimators were much less pronounced than reproducibility differences between individual subjects.
 
 As a representative for the reproducibility of graph metrics, we investigated betweenness. Here, $\nAC$ offered the highest reproducibility (average (sd): 0.58 (0.18)) and $\nlCA$ improved markedly over $\nCA$ (0.41 (0.15) vs 0.30 (0.17)). This is contrast to another study that found no effect of aggregation method (region mean time series versus region median  versus 1st eigenvariate of the region) on the reproducibility of graph metrics~\citep{braun_testretest_2012} (although in that study sessions were weeks apart).

\subsection{Discriminability}
        Estimators $\nCA,\nAC,\nlCA$ showed similar values for discriminability, with slightly lower identification rate and intra-subject to inter-subject distribution separation for $\nAC$ than the two others, and slightly lower intra-inter separation for $\nlCA$ than $\nCA$. This suggests that the improved robustness to region size and intra-correlation effects of $\nlCA$ does not result in a sizeable impact on discriminative ability, although this warrants further evaluation.
 
\subsection{Limitations}
        Our signal model, and therefore the derived estimators, is a trade-off between model realism and tractability of the analysis of estimator properties. This comes with important limitations.
        
        First, assuming stationarity and additivity of the local noise fails to capture effects like system instability due to B0 inhomogeneity, RF power variations, or gradient fluctuations~\citep{lazar_noise_2008, greve_survey_2013, liu_noise_2016}. Independently of the model, note that effects such as drift are mitigated by using wavelet coefficient time series as we did in this study, and that such instabilities explain proportionally less of the noise variance than thermal noise at high field~\citep{greve_novel_2011}.
        
        Second, motion effects, and in particular differential long-vs short-range effects on correlations~\citep{van_dijk_intrinsic_2010, yan_comprehensive_2013}, were not studied, and their interplay with the spatial bias exhibited by estimator $\nCA$ in Figure~\ref{fig:braingraph_diffs} was not examined. 
        
        Third, our new estimators come with the added burden of choosing hyperparameters such as neighbourhood size. These are currently selected empirically, and no systematic sensitivity analysis has been performed.
        
        Despite these limitations, we believe our empirical tests served to bridge the gap towards applicability, since our model yielded at least an estimator, $\nlCA$, with useful properties for use in neuroimaging - namely, reduced dependency to region size and low intra-correlation, and improved reproducibility of graph metrics.

\section{Acknowledgements}

This work pas partly funded by French National Research Agency project ANR-20-NEUC-0003-02. Rat data was acquired at the IRMaGe MRI facility, partly funded by the French program Investissement d’Avenir run by the French National Research Agency, grant Infrastructure d’avenir en Biologie Santé ANR-11-INBS-0006.
Human data were provided by the Human Connectome Project, WU-Minn Consortium (Principal Investigators: David Van Essen and Kamil Ugurbil; 1U54MH091657) funded by the 16 NIH Institutes and Centers that support the NIH Blueprint for Neuroscience Research; and by the McDonnell Center for Systems Neuroscience at Washington University. Codes are available at \url{https://gitlab.inria.fr/q-func/ireco4fmri/}, datasets are available at \url{https://dx.doi.org/10.5281/zenodo.7254133}. 
	
	\bibliographystyle{model2-names}
	\bibliography{biblio,referencesFromZotero}

\begin{thebibliography}{57}
\expandafter\ifx\csname natexlab\endcsname\relax\def\natexlab#1{#1}\fi
\expandafter\ifx\csname url\endcsname\relax
  \def\url#1{\texttt{#1}}\fi
\expandafter\ifx\csname urlprefix\endcsname\relax\def\urlprefix{URL }\fi
\providecommand{\eprint}[2][]{\url{#2}}
\providecommand{\bibinfo}[2]{#2}
\ifx\xfnm\relax \def\xfnm[#1]{\unskip,\space#1}\fi
\bibitem[{Achard et~al.(2011)Achard, Coeurjolly, Marcillaud and
  Richiardi}]{ssp.2011.1}
\bibinfo{author}{Achard, S.}, \bibinfo{author}{Coeurjolly, J.F.},
  \bibinfo{author}{Marcillaud, R.}, \bibinfo{author}{Richiardi, J.},
  \bibinfo{year}{2011}.
\newblock \bibinfo{title}{{fMRI} functional connectivity estimators robust to
  region size bias}, in: \bibinfo{booktitle}{IEEE Workshop on Statistical
  Signal Processing, SSP2011}, \bibinfo{address}{Nice, France}. pp.
  \bibinfo{pages}{813--816}.
\bibitem[{Achard et~al.(2006)Achard, Salvador, Whitcher, Suckling and
  Bullmore}]{achard.2006.1}
\bibinfo{author}{Achard, S.}, \bibinfo{author}{Salvador, R.},
  \bibinfo{author}{Whitcher, B.}, \bibinfo{author}{Suckling, J.},
  \bibinfo{author}{Bullmore, E.}, \bibinfo{year}{2006}.
\newblock \bibinfo{title}{A resilient, low-frequency, small-world human brain
  functional network with highly connected association cortical hubs}.
\newblock \bibinfo{journal}{Journal of Neuroscience} \bibinfo{volume}{26},
  \bibinfo{pages}{63--72}.
\bibitem[{Afyouni et~al.(2019)Afyouni, Smith and
  Nichols}]{afyouni_effective_2019}
\bibinfo{author}{Afyouni, S.}, \bibinfo{author}{Smith, S.M.},
  \bibinfo{author}{Nichols, T.E.}, \bibinfo{year}{2019}.
\newblock \bibinfo{title}{Effective degrees of freedom of the {Pearson}'s
  correlation coefficient under autocorrelation}.
\newblock \bibinfo{journal}{NeuroImage} \bibinfo{volume}{199},
  \bibinfo{pages}{609--625}.
\bibitem[{Alexander-Bloch et~al.(2010)Alexander-Bloch, Gogtay, Meunier, Birn,
  Clasen, Lalonde, Lenroot, Giedd and
  Bullmore}]{alexander-bloch_disrupted_2010}
\bibinfo{author}{Alexander-Bloch, A.F.}, \bibinfo{author}{Gogtay, N.},
  \bibinfo{author}{Meunier, D.}, \bibinfo{author}{Birn, R.},
  \bibinfo{author}{Clasen, L.}, \bibinfo{author}{Lalonde, F.},
  \bibinfo{author}{Lenroot, R.}, \bibinfo{author}{Giedd, J.},
  \bibinfo{author}{Bullmore, E.T.}, \bibinfo{year}{2010}.
\newblock \bibinfo{title}{Disrupted {Modularity} and {Local} {Connectivity} of
  {Brain} {Functional} {Networks} in {Childhood}-{Onset} {Schizophrenia}}.
\newblock \bibinfo{journal}{Frontiers in Systems Neuroscience}
  \bibinfo{volume}{4}.
\bibitem[{Becq et~al.(2020a)Becq, Barbier and
  Achard}]{becq_10.1088/1741-2552/ab9fec}
\bibinfo{author}{Becq, G.G.J.P.C.}, \bibinfo{author}{Barbier, E.},
  \bibinfo{author}{Achard, S.}, \bibinfo{year}{2020}a.
\newblock \bibinfo{title}{Brain networks of rats under anesthesia using
  resting-state fmri: comparison with dead rats, random noise and generative
  models of networks}.
\newblock \bibinfo{journal}{Journal of Neural Engineering} .
\bibitem[{Becq et~al.(2020b)Becq, Habet, Collomb, Faucher, Delon-Martin,
  Coizet, Achard and Barbier}]{guillaume2020functional}
\bibinfo{author}{Becq, G.J.P.}, \bibinfo{author}{Habet, T.},
  \bibinfo{author}{Collomb, N.}, \bibinfo{author}{Faucher, M.},
  \bibinfo{author}{Delon-Martin, C.}, \bibinfo{author}{Coizet, V.},
  \bibinfo{author}{Achard, S.}, \bibinfo{author}{Barbier, E.L.},
  \bibinfo{year}{2020}b.
\newblock \bibinfo{title}{Functional connectivity is preserved but reorganized
  across several anesthetic regimes}.
\newblock \bibinfo{journal}{NeuroImage} \bibinfo{volume}{219},
  \bibinfo{pages}{116945}.
\bibitem[{Bergholm et~al.(2010)Bergholm, Adler and Parmryd}]{bergholm.2010.1}
\bibinfo{author}{Bergholm, F.}, \bibinfo{author}{Adler, J.},
  \bibinfo{author}{Parmryd, I.}, \bibinfo{year}{2010}.
\newblock \bibinfo{title}{Analysis of bias in the apparent correlation
  coefficient between image pairs corrupted by severe noise}.
\newblock \bibinfo{journal}{J Math Imaging Vis} \bibinfo{volume}{37},
  \bibinfo{pages}{204--219}.
\bibitem[{Bolt et~al.(2017)Bolt, Nomi, Rubinov and Uddin}]{Bolt2017CA}
\bibinfo{author}{Bolt, T.}, \bibinfo{author}{Nomi, J.S.},
  \bibinfo{author}{Rubinov, M.}, \bibinfo{author}{Uddin, L.Q.},
  \bibinfo{year}{2017}.
\newblock \bibinfo{title}{Correspondence between evoked and intrinsic
  functional brain network configurations}.
\newblock \bibinfo{journal}{Human Brain Mapping} \bibinfo{volume}{38},
  \bibinfo{pages}{1992--2007}.
\newblock \eprint{https://onlinelibrary.wiley.com/doi/pdf/10.1002/hbm.23500}.
\bibitem[{Braun et~al.(2012)Braun, Plichta, Esslinger, Sauer, Haddad, Grimm,
  Mier, Mohnke, Heinz, Erk, Walter, Seiferth, Kirsch and
  Meyer-Lindenberg}]{braun_testretest_2012}
\bibinfo{author}{Braun, U.}, \bibinfo{author}{Plichta, M.M.},
  \bibinfo{author}{Esslinger, C.}, \bibinfo{author}{Sauer, C.},
  \bibinfo{author}{Haddad, L.}, \bibinfo{author}{Grimm, O.},
  \bibinfo{author}{Mier, D.}, \bibinfo{author}{Mohnke, S.},
  \bibinfo{author}{Heinz, A.}, \bibinfo{author}{Erk, S.},
  \bibinfo{author}{Walter, H.}, \bibinfo{author}{Seiferth, N.},
  \bibinfo{author}{Kirsch, P.}, \bibinfo{author}{Meyer-Lindenberg, A.},
  \bibinfo{year}{2012}.
\newblock \bibinfo{title}{Test–retest reliability of resting-state
  connectivity network characteristics using {fMRI} and graph theoretical
  measures}.
\newblock \bibinfo{journal}{NeuroImage} \bibinfo{volume}{59},
  \bibinfo{pages}{1404--1412}.
\bibitem[{Büchel and Friston(1997)}]{buchel_modulation_1997}
\bibinfo{author}{Büchel, C.}, \bibinfo{author}{Friston, K.J.},
  \bibinfo{year}{1997}.
\newblock \bibinfo{title}{Modulation of connectivity in visual pathways by
  attention: cortical interactions evaluated with structural equation modelling
  and {fMRI}.}
\newblock \bibinfo{journal}{Cerebral Cortex} \bibinfo{volume}{7},
  \bibinfo{pages}{768--778}.
\bibitem[{Caballero-Gaudes and Reynolds(2017)}]{caballero-gaudes_methods_2017}
\bibinfo{author}{Caballero-Gaudes, C.}, \bibinfo{author}{Reynolds, R.C.},
  \bibinfo{year}{2017}.
\newblock \bibinfo{title}{Methods for cleaning the {BOLD} {fMRI} signal}.
\newblock \bibinfo{journal}{NeuroImage} \bibinfo{volume}{154},
  \bibinfo{pages}{128--149}.
\bibitem[{Cao et~al.(2019)Cao, McEwen, Forsyth, Gee, Bearden, Addington,
  Goodyear, Cadenhead, Mirzakhanian, Cornblatt, Carrión, Mathalon, McGlashan,
  Perkins, Belger, Seidman, Thermenos, Tsuang, van Erp, Walker, Hamann,
  Anticevic, Woods and Cannon}]{cao_toward_2019}
\bibinfo{author}{Cao, H.}, \bibinfo{author}{McEwen, S.C.},
  \bibinfo{author}{Forsyth, J.K.}, \bibinfo{author}{Gee, D.G.},
  \bibinfo{author}{Bearden, C.E.}, \bibinfo{author}{Addington, J.},
  \bibinfo{author}{Goodyear, B.}, \bibinfo{author}{Cadenhead, K.S.},
  \bibinfo{author}{Mirzakhanian, H.}, \bibinfo{author}{Cornblatt, B.A.},
  \bibinfo{author}{Carrión, R.E.}, \bibinfo{author}{Mathalon, D.H.},
  \bibinfo{author}{McGlashan, T.H.}, \bibinfo{author}{Perkins, D.O.},
  \bibinfo{author}{Belger, A.}, \bibinfo{author}{Seidman, L.J.},
  \bibinfo{author}{Thermenos, H.}, \bibinfo{author}{Tsuang, M.T.},
  \bibinfo{author}{van Erp, T.G.M.}, \bibinfo{author}{Walker, E.F.},
  \bibinfo{author}{Hamann, S.}, \bibinfo{author}{Anticevic, A.},
  \bibinfo{author}{Woods, S.W.}, \bibinfo{author}{Cannon, T.D.},
  \bibinfo{year}{2019}.
\newblock \bibinfo{title}{Toward {Leveraging} {Human} {Connectomic} {Data} in
  {Large} {Consortia}: {Generalizability} of {fMRI}-{Based} {Brain} {Graphs}
  {Across} {Sites}, {Sessions}, and {Paradigms}}.
\newblock \bibinfo{journal}{Cerebral Cortex} \bibinfo{volume}{29},
  \bibinfo{pages}{1263--1279}.
\bibitem[{Chen and Glover(2015)}]{chen_bold_2015}
\bibinfo{author}{Chen, J.E.}, \bibinfo{author}{Glover, G.H.},
  \bibinfo{year}{2015}.
\newblock \bibinfo{title}{{BOLD} fractional contribution to resting-state
  functional connectivity above 0.1 {Hz}}.
\newblock \bibinfo{journal}{NeuroImage} \bibinfo{volume}{107},
  \bibinfo{pages}{207--218}.
\bibitem[{Clifford et~al.(1989)Clifford, Richardson and
  Hemon}]{clifford.1989.1}
\bibinfo{author}{Clifford, P.}, \bibinfo{author}{Richardson, S.},
  \bibinfo{author}{Hemon, D.}, \bibinfo{year}{1989}.
\newblock \bibinfo{title}{Assessing the significance of the correlation between
  two spatial processes}.
\newblock \bibinfo{journal}{Biometrics} \bibinfo{volume}{45},
  \bibinfo{pages}{123--134}.
\bibitem[{Cordes et~al.(2001)Cordes, Haughton, Arfanakis, Carew, Turski,
  Moritz, Quigley and Meyerand}]{cordes_frequencies_2001}
\bibinfo{author}{Cordes, D.}, \bibinfo{author}{Haughton, V.M.},
  \bibinfo{author}{Arfanakis, K.}, \bibinfo{author}{Carew, J.D.},
  \bibinfo{author}{Turski, P.A.}, \bibinfo{author}{Moritz, C.H.},
  \bibinfo{author}{Quigley, M.A.}, \bibinfo{author}{Meyerand, M.E.},
  \bibinfo{year}{2001}.
\newblock \bibinfo{title}{Frequencies {Contributing} to {Functional}
  {Connectivity} in the {Cerebral} {Cortex} in “{Resting}-state” {Data}}.
\newblock \bibinfo{journal}{American Journal of Neuroradiology}
  \bibinfo{volume}{22}, \bibinfo{pages}{1326--1333}.
\newblock \bibinfo{note}{Publisher: American Journal of Neuroradiology Section:
  BRAIN}.
\bibitem[{Dadi et~al.(2019)Dadi, Rahim, Abraham, Chyzhyk, Milham, Thirion and
  Varoquaux}]{dadi_benchmarking_2019}
\bibinfo{author}{Dadi, K.}, \bibinfo{author}{Rahim, M.},
  \bibinfo{author}{Abraham, A.}, \bibinfo{author}{Chyzhyk, D.},
  \bibinfo{author}{Milham, M.}, \bibinfo{author}{Thirion, B.},
  \bibinfo{author}{Varoquaux, G.}, \bibinfo{year}{2019}.
\newblock \bibinfo{title}{Benchmarking functional connectome-based predictive
  models for resting-state {fMRI}}.
\newblock \bibinfo{journal}{NeuroImage} \bibinfo{volume}{192},
  \bibinfo{pages}{115--134}.
\bibitem[{Deco et~al.(2014)Deco, Ponce-Alvarez, Hagmann, Romani, Mantini and
  Corbetta}]{deco_how_2014}
\bibinfo{author}{Deco, G.}, \bibinfo{author}{Ponce-Alvarez, A.},
  \bibinfo{author}{Hagmann, P.}, \bibinfo{author}{Romani, G.L.},
  \bibinfo{author}{Mantini, D.}, \bibinfo{author}{Corbetta, M.},
  \bibinfo{year}{2014}.
\newblock \bibinfo{title}{How {Local} {Excitation}-{Inhibition} {Ratio}
  {Impacts} the {Whole} {Brain} {Dynamics}}.
\newblock \bibinfo{journal}{Journal of Neuroscience} \bibinfo{volume}{34},
  \bibinfo{pages}{7886--7898}.
\bibitem[{Donner and Eliasziw(1991)}]{donner.1991.1}
\bibinfo{author}{Donner, A.}, \bibinfo{author}{Eliasziw, M.},
  \bibinfo{year}{1991}.
\newblock \bibinfo{title}{Methodology for inferences concerning familial
  correlations: a review}.
\newblock \bibinfo{journal}{J Clin Epidemio} \bibinfo{volume}{44},
  \bibinfo{pages}{449--455}.
\bibitem[{Eickhoff et~al.(2015)Eickhoff, Thirion, Varoquaux and
  Bzdok}]{eickhoff_connectivitybased_2015}
\bibinfo{author}{Eickhoff, S.B.}, \bibinfo{author}{Thirion, B.},
  \bibinfo{author}{Varoquaux, G.}, \bibinfo{author}{Bzdok, D.},
  \bibinfo{year}{2015}.
\newblock \bibinfo{title}{Connectivity‐based parcellation: {Critique} and
  implications}.
\newblock \bibinfo{journal}{Human Brain Mapping} \bibinfo{volume}{36},
  \bibinfo{pages}{4771--4792}.
\bibitem[{Figueroa-Jimenez et~al.(2021)Figueroa-Jimenez, Cañete-Massé,
  Carbó-Carreté, Zarabozo-Hurtado, Peró-Cebollero, Salazar-Estrada and
  Guàrdia-Olmos}]{Figueroa2021CA}
\bibinfo{author}{Figueroa-Jimenez, M.D.}, \bibinfo{author}{Cañete-Massé, C.},
  \bibinfo{author}{Carbó-Carreté, M.}, \bibinfo{author}{Zarabozo-Hurtado,
  D.}, \bibinfo{author}{Peró-Cebollero, M.}, \bibinfo{author}{Salazar-Estrada,
  J.G.}, \bibinfo{author}{Guàrdia-Olmos, J.}, \bibinfo{year}{2021}.
\newblock \bibinfo{title}{Resting-state default mode network connectivity in
  young individuals with down syndrome}.
\newblock \bibinfo{journal}{Brain and Behavior} \bibinfo{volume}{11},
  \bibinfo{pages}{e01905}.
\newblock \eprint{https://onlinelibrary.wiley.com/doi/pdf/10.1002/brb3.1905}.
\bibitem[{Finn et~al.(2015)Finn, Shen, Scheinost, Rosenberg, Huang, Chun,
  Papademetris and Constable}]{finn_functional_2015}
\bibinfo{author}{Finn, E.S.}, \bibinfo{author}{Shen, X.},
  \bibinfo{author}{Scheinost, D.}, \bibinfo{author}{Rosenberg, M.D.},
  \bibinfo{author}{Huang, J.}, \bibinfo{author}{Chun, M.M.},
  \bibinfo{author}{Papademetris, X.}, \bibinfo{author}{Constable, R.T.},
  \bibinfo{year}{2015}.
\newblock \bibinfo{title}{Functional connectome fingerprinting: identifying
  individuals using patterns of brain connectivity}.
\newblock \bibinfo{journal}{Nature Neuroscience} \bibinfo{volume}{18},
  \bibinfo{pages}{1664--1671}.
\bibitem[{Glasser et~al.(2013)Glasser, Sotiropoulos, Wilson, Coalson, Fischl,
  Andersson, Xu, Jbabdi, Webster, Polimeni, Van~Essen and
  Jenkinson}]{glasser_minimal_2013}
\bibinfo{author}{Glasser, M.F.}, \bibinfo{author}{Sotiropoulos, S.N.},
  \bibinfo{author}{Wilson, J.A.}, \bibinfo{author}{Coalson, T.S.},
  \bibinfo{author}{Fischl, B.}, \bibinfo{author}{Andersson, J.L.},
  \bibinfo{author}{Xu, J.}, \bibinfo{author}{Jbabdi, S.},
  \bibinfo{author}{Webster, M.}, \bibinfo{author}{Polimeni, J.R.},
  \bibinfo{author}{Van~Essen, D.C.}, \bibinfo{author}{Jenkinson, M.},
  \bibinfo{year}{2013}.
\newblock \bibinfo{title}{The minimal preprocessing pipelines for the {Human}
  {Connectome} {Project}}.
\newblock \bibinfo{journal}{NeuroImage} \bibinfo{volume}{80},
  \bibinfo{pages}{105--124}.
\bibitem[{Greve et~al.(2013)Greve, Brown, Mueller, Glover, Liu and {Function
  Biomedical Research Network}}]{greve_survey_2013}
\bibinfo{author}{Greve, D.N.}, \bibinfo{author}{Brown, G.G.},
  \bibinfo{author}{Mueller, B.A.}, \bibinfo{author}{Glover, G.},
  \bibinfo{author}{Liu, T.T.}, \bibinfo{author}{{Function Biomedical Research
  Network}}, \bibinfo{year}{2013}.
\newblock \bibinfo{title}{A {Survey} of the {Sources} of {Noise} in {fMRI}}.
\newblock \bibinfo{journal}{Psychometrika} \bibinfo{volume}{78},
  \bibinfo{pages}{396--416}.
\bibitem[{Greve et~al.(2011)Greve, Mueller, Liu, Turner, Voyvodic, Yetter,
  Diaz, McCarthy, Wallace, Roach, Ford, Mathalon, Calhoun, Wible, Brown, Potkin
  and Glover}]{greve_novel_2011}
\bibinfo{author}{Greve, D.N.}, \bibinfo{author}{Mueller, B.A.},
  \bibinfo{author}{Liu, T.}, \bibinfo{author}{Turner, J.A.},
  \bibinfo{author}{Voyvodic, J.}, \bibinfo{author}{Yetter, E.},
  \bibinfo{author}{Diaz, M.}, \bibinfo{author}{McCarthy, G.},
  \bibinfo{author}{Wallace, S.}, \bibinfo{author}{Roach, B.J.},
  \bibinfo{author}{Ford, J.M.}, \bibinfo{author}{Mathalon, D.H.},
  \bibinfo{author}{Calhoun, V.D.}, \bibinfo{author}{Wible, C.G.},
  \bibinfo{author}{Brown, G.G.}, \bibinfo{author}{Potkin, S.G.},
  \bibinfo{author}{Glover, G.}, \bibinfo{year}{2011}.
\newblock \bibinfo{title}{A novel method for quantifying scanner instability in
  {fMRI}: {Quantifying} {Scanner} {Instability} in {fMRI}}.
\newblock \bibinfo{journal}{Magnetic Resonance in Medicine}
  \bibinfo{volume}{65}, \bibinfo{pages}{1053--1061}.
\bibitem[{Gunst(1995)}]{gunst.1995.1}
\bibinfo{author}{Gunst, R.F.}, \bibinfo{year}{1995}.
\newblock \bibinfo{title}{Estimating spatial correlations from spatial-temporal
  meteorological data}.
\newblock \bibinfo{journal}{Journal of climate} \bibinfo{volume}{8},
  \bibinfo{pages}{2454--69}.
\bibitem[{Jiang and Zuo(2016)}]{jiang_regional_2016}
\bibinfo{author}{Jiang, L.}, \bibinfo{author}{Zuo, X.N.}, \bibinfo{year}{2016}.
\newblock \bibinfo{title}{Regional {Homogeneity}: {A} {Multimodal},
  {Multiscale} {Neuroimaging} {Marker} of the {Human} {Connectome}}.
\newblock \bibinfo{journal}{The Neuroscientist} \bibinfo{volume}{22},
  \bibinfo{pages}{486--505}.
\bibitem[{Jo et~al.(2010)Jo, Saad, Simmons, Milbury and Cox}]{jo_mapping_2010}
\bibinfo{author}{Jo, H.J.}, \bibinfo{author}{Saad, Z.S.},
  \bibinfo{author}{Simmons, W.K.}, \bibinfo{author}{Milbury, L.A.},
  \bibinfo{author}{Cox, R.W.}, \bibinfo{year}{2010}.
\newblock \bibinfo{title}{Mapping sources of correlation in resting state
  {FMRI}, with artifact detection and removal}.
\newblock \bibinfo{journal}{NeuroImage} \bibinfo{volume}{52},
  \bibinfo{pages}{571--582}.
\bibitem[{Köhler et~al.(1998)Köhler, McIntosh, Moscovitch and
  Winocur}]{kohler_functional_1998}
\bibinfo{author}{Köhler, S.}, \bibinfo{author}{McIntosh, A.R.},
  \bibinfo{author}{Moscovitch, M.}, \bibinfo{author}{Winocur, G.},
  \bibinfo{year}{1998}.
\newblock \bibinfo{title}{Functional interactions between the medial temporal
  lobes and posterior neocortex related to episodic memory retrieval.}
\newblock \bibinfo{journal}{Cerebral Cortex} \bibinfo{volume}{8},
  \bibinfo{pages}{451--461}.
\bibitem[{Lazar(2008)}]{lazar_noise_2008}
\bibinfo{author}{Lazar, N.A.}, \bibinfo{year}{2008}.
\newblock \bibinfo{title}{Noise and {Data} {Preprocessing}}, in:
  \bibinfo{booktitle}{The {Statistical} {Analysis} of {Functional} {MRI}
  {Data}}. \bibinfo{publisher}{Springer New York}, \bibinfo{address}{New York,
  NY}, pp. \bibinfo{pages}{37--51}.
\newblock \bibinfo{note}{ISSN: 1431-8776 Series Title: Statistics for Biology
  and Health}.
\bibitem[{Liebhold and Sharov(1998)}]{liebhold.1998.1}
\bibinfo{author}{Liebhold, A.}, \bibinfo{author}{Sharov, A.},
  \bibinfo{year}{1998}.
\newblock \bibinfo{title}{Testing for correlation in the presence of spatial
  autocorrelation in insect count data}.
\newblock \bibinfo{journal}{Population and Community Ecology for Insect
  Management and Conservation} , \bibinfo{pages}{1--8}.
\bibitem[{Lin(1989)}]{lin_concordance_1989}
\bibinfo{author}{Lin, L.I.K.}, \bibinfo{year}{1989}.
\newblock \bibinfo{title}{A {Concordance} {Correlation} {Coefficient} to
  {Evaluate} {Reproducibility}}.
\newblock \bibinfo{journal}{Biometrics} \bibinfo{volume}{45},
  \bibinfo{pages}{255--268}.
\newblock \bibinfo{note}{Publisher: [Wiley, International Biometric Society]}.
\bibitem[{Liu(2016)}]{liu_noise_2016}
\bibinfo{author}{Liu, T.T.}, \bibinfo{year}{2016}.
\newblock \bibinfo{title}{Noise contributions to the {fMRI} signal: {An}
  overview}.
\newblock \bibinfo{journal}{NeuroImage} \bibinfo{volume}{143},
  \bibinfo{pages}{141--151}.
\bibitem[{Moulines et~al.(2007)Moulines, Roueff and
  Taqqu}]{Moulines07SpectralDensity}
\bibinfo{author}{Moulines, E.}, \bibinfo{author}{Roueff, F.},
  \bibinfo{author}{Taqqu, M.S.}, \bibinfo{year}{2007}.
\newblock \bibinfo{title}{On the spectral density of the wavelet coefficients
  of long-memory time series with application to the log-regression estimation
  of the memory parameter}.
\newblock \bibinfo{journal}{Journal of Time Series Analysis}
  \bibinfo{volume}{28}, \bibinfo{pages}{155--187}.
\bibitem[{Murphy et~al.(2013)Murphy, Birn and
  Bandettini}]{murphy_resting-state_2013}
\bibinfo{author}{Murphy, K.}, \bibinfo{author}{Birn, R.M.},
  \bibinfo{author}{Bandettini, P.A.}, \bibinfo{year}{2013}.
\newblock \bibinfo{title}{Resting-state {fMRI} confounds and cleanup}.
\newblock \bibinfo{journal}{NeuroImage} \bibinfo{volume}{80},
  \bibinfo{pages}{349--359}.
\bibitem[{Ng et~al.(2016)Ng, Varoquaux, Poline, Greicius and
  Thirion}]{ng_transport_2016}
\bibinfo{author}{Ng, B.}, \bibinfo{author}{Varoquaux, G.},
  \bibinfo{author}{Poline, J.B.}, \bibinfo{author}{Greicius, M.},
  \bibinfo{author}{Thirion, B.}, \bibinfo{year}{2016}.
\newblock \bibinfo{title}{Transport on {Riemannian} {Manifold} for
  {Connectivity}-{Based} {Brain} {Decoding}}.
\newblock \bibinfo{journal}{IEEE Transactions on Medical Imaging}
  \bibinfo{volume}{35}, \bibinfo{pages}{208--216}.
\bibitem[{Ogawa(2021)}]{Akitoshi2021CA}
\bibinfo{author}{Ogawa, A.}, \bibinfo{year}{2021}.
\newblock \bibinfo{title}{Time-varying measures of cerebral network centrality
  correlate with visual saliency during movie watching}.
\newblock \bibinfo{journal}{Brain and Behavior} \bibinfo{volume}{11},
  \bibinfo{pages}{e2334}.
\newblock \eprint{https://onlinelibrary.wiley.com/doi/pdf/10.1002/brb3.2334}.
\bibitem[{Ostroff(1993)}]{ostroff1993comparing}
\bibinfo{author}{Ostroff, C.}, \bibinfo{year}{1993}.
\newblock \bibinfo{title}{Comparing correlations based on individual-level and
  aggregated data.}
\newblock \bibinfo{journal}{Journal of Applied Psychology}
  \bibinfo{volume}{78}, \bibinfo{pages}{569}.
\bibitem[{Pawela et~al.(2008)Pawela, Biswal, Cho, Kao, Li, Jones, Schulte,
  Matloub, Hudetz and Hyde}]{Pawela2008}
\bibinfo{author}{Pawela, C.P.}, \bibinfo{author}{Biswal, B.B.},
  \bibinfo{author}{Cho, Y.R.}, \bibinfo{author}{Kao, D.S.},
  \bibinfo{author}{Li, R.}, \bibinfo{author}{Jones, S.R.},
  \bibinfo{author}{Schulte, M.L.}, \bibinfo{author}{Matloub, H.S.},
  \bibinfo{author}{Hudetz, A.G.}, \bibinfo{author}{Hyde, J.S.},
  \bibinfo{year}{2008}.
\newblock \bibinfo{title}{Resting-state functional connectivity of the rat
  brain}.
\newblock \bibinfo{journal}{Magnetic Resonance in Medicine}
  \bibinfo{volume}{59}, \bibinfo{pages}{1021--1029}.
\bibitem[{Petersen et~al.(2016)Petersen, Zhao, Carmichael and
  Müller}]{petersen_quantifying_2016}
\bibinfo{author}{Petersen, A.}, \bibinfo{author}{Zhao, J.},
  \bibinfo{author}{Carmichael, O.}, \bibinfo{author}{Müller, H.G.},
  \bibinfo{year}{2016}.
\newblock \bibinfo{title}{Quantifying {Individual} {Brain} {Connectivity} with
  {Functional} {Principal} {Component} {Analysis} for {Networks}}.
\newblock \bibinfo{journal}{Brain Connectivity} \bibinfo{volume}{6},
  \bibinfo{pages}{540--547}.
\bibitem[{Poline et~al.(1997)Poline, Worsley, Evans and
  Friston}]{poline_combining_1997}
\bibinfo{author}{Poline, J.B.}, \bibinfo{author}{Worsley, K.J.},
  \bibinfo{author}{Evans, A.C.}, \bibinfo{author}{Friston, K.J.},
  \bibinfo{year}{1997}.
\newblock \bibinfo{title}{Combining spatial extent and peak intensity to test
  for activations in functional imaging}.
\newblock \bibinfo{journal}{NeuroImage} \bibinfo{volume}{5},
  \bibinfo{pages}{83--96}.
\bibitem[{Richiardi et~al.(2013)Richiardi, Achard, Bunke and Van
  De~Ville}]{richiardi_machine_2013}
\bibinfo{author}{Richiardi, J.}, \bibinfo{author}{Achard, S.},
  \bibinfo{author}{Bunke, H.}, \bibinfo{author}{Van De~Ville, D.},
  \bibinfo{year}{2013}.
\newblock \bibinfo{title}{Machine {Learning} with {Brain} {Graphs}:
  {Predictive} {Modeling} {Approaches} for {Functional} {Imaging} in {Systems}
  {Neuroscience}}.
\newblock \bibinfo{journal}{IEEE Signal Processing Magazine}
  \bibinfo{volume}{30}, \bibinfo{pages}{58--70}.
\bibitem[{Rosner et~al.(1977)Rosner, Donner and Hennekens}]{rosner.1977.1}
\bibinfo{author}{Rosner, B.}, \bibinfo{author}{Donner, A.},
  \bibinfo{author}{Hennekens, C.}, \bibinfo{year}{1977}.
\newblock \bibinfo{title}{Estimation of interclass correlation from familial
  data}.
\newblock \bibinfo{journal}{Applied statistics} \bibinfo{volume}{26},
  \bibinfo{pages}{179--187}.
\bibitem[{Salvador et~al.(2005)Salvador, Suckling, Schwarzbauer and
  Bullmore}]{salvador_undirected_2005}
\bibinfo{author}{Salvador, R.}, \bibinfo{author}{Suckling, J.},
  \bibinfo{author}{Schwarzbauer, C.}, \bibinfo{author}{Bullmore, E.},
  \bibinfo{year}{2005}.
\newblock \bibinfo{title}{Undirected graphs of frequency-dependent functional
  connectivity in whole brain networks}.
\newblock \bibinfo{journal}{Philosophical Transactions of the Royal Society B:
  Biological Sciences} \bibinfo{volume}{360}, \bibinfo{pages}{937--946}.
\bibitem[{Stanley et~al.(2013)Stanley, Moussa, Paolini, Lyday, Burdette and
  Laurienti}]{stanley_defining_2013}
\bibinfo{author}{Stanley, M.L.}, \bibinfo{author}{Moussa, M.N.},
  \bibinfo{author}{Paolini, B.M.}, \bibinfo{author}{Lyday, R.G.},
  \bibinfo{author}{Burdette, J.H.}, \bibinfo{author}{Laurienti, P.J.},
  \bibinfo{year}{2013}.
\newblock \bibinfo{title}{Defining nodes in complex brain networks}.
\newblock \bibinfo{journal}{Frontiers in Computational Neuroscience}
  \bibinfo{volume}{7}.
\bibitem[{Student(1914)}]{student.1914.1}
\bibinfo{author}{Student}, \bibinfo{year}{1914}.
\newblock \bibinfo{title}{The elimination of spurious correlation due to
  position in time and space}.
\newblock \bibinfo{journal}{Biometrika} \bibinfo{volume}{10},
  \bibinfo{pages}{179--181}.
\bibitem[{Termenon et~al.(2016)Termenon, Jaillard, Delon-Martin and
  Achard}]{termenon2016reliability}
\bibinfo{author}{Termenon, M.}, \bibinfo{author}{Jaillard, A.},
  \bibinfo{author}{Delon-Martin, C.}, \bibinfo{author}{Achard, S.},
  \bibinfo{year}{2016}.
\newblock \bibinfo{title}{Reliability of graph analysis of resting state {fMRI}
  using test-retest dataset from the human connectome project}.
\newblock \bibinfo{journal}{Neuroimage} \bibinfo{volume}{142},
  \bibinfo{pages}{172--187}.
\bibitem[{Triana et~al.(2020)Triana, Glerean, Saram{\"a}ki and
  Korhonen}]{triana2020effects}
\bibinfo{author}{Triana, A.M.}, \bibinfo{author}{Glerean, E.},
  \bibinfo{author}{Saram{\"a}ki, J.}, \bibinfo{author}{Korhonen, O.},
  \bibinfo{year}{2020}.
\newblock \bibinfo{title}{Effects of spatial smoothing on group-level
  differences in functional brain networks}.
\newblock \bibinfo{journal}{Network Neuroscience} \bibinfo{volume}{4},
  \bibinfo{pages}{556--574}.
\bibitem[{Van~Dijk et~al.(2010)Van~Dijk, Hedden, Venkataraman, Evans, Lazar and
  Buckner}]{van_dijk_intrinsic_2010}
\bibinfo{author}{Van~Dijk, K.R.A.}, \bibinfo{author}{Hedden, T.},
  \bibinfo{author}{Venkataraman, A.}, \bibinfo{author}{Evans, K.C.},
  \bibinfo{author}{Lazar, S.W.}, \bibinfo{author}{Buckner, R.L.},
  \bibinfo{year}{2010}.
\newblock \bibinfo{title}{Intrinsic {Functional} {Connectivity} {As} a {Tool}
  {For} {Human} {Connectomics}: {Theory}, {Properties}, and {Optimization}}.
\newblock \bibinfo{journal}{Journal of Neurophysiology} \bibinfo{volume}{103},
  \bibinfo{pages}{297--321}.
\bibitem[{Whitlow et~al.(2011)Whitlow, Casanova and
  Maldjian}]{whitlow_effect_2011}
\bibinfo{author}{Whitlow, C.T.}, \bibinfo{author}{Casanova, R.},
  \bibinfo{author}{Maldjian, J.A.}, \bibinfo{year}{2011}.
\newblock \bibinfo{title}{Effect of {Resting}-{State} {Functional} {MR}
  {Imaging} {Duration} on {Stability} of {Graph} {Theory} {Metrics} of {Brain}
  {Network} {Connectivity}}.
\newblock \bibinfo{journal}{Radiology} \bibinfo{volume}{259},
  \bibinfo{pages}{516--524}.
\bibitem[{Wilson(2019)}]{wilson_harmonic_2019}
\bibinfo{author}{Wilson, D.J.}, \bibinfo{year}{2019}.
\newblock \bibinfo{title}{The harmonic mean \textit{p} -value for combining
  dependent tests}.
\newblock \bibinfo{journal}{Proceedings of the National Academy of Sciences}
  \bibinfo{volume}{116}, \bibinfo{pages}{1195--1200}.
\bibitem[{Worsley et~al.(1992)Worsley, Evans, Marrett and
  Neelin}]{worsley_three-dimensional_1992}
\bibinfo{author}{Worsley, K.J.}, \bibinfo{author}{Evans, A.C.},
  \bibinfo{author}{Marrett, S.}, \bibinfo{author}{Neelin, P.},
  \bibinfo{year}{1992}.
\newblock \bibinfo{title}{A {Three}-{Dimensional} {Statistical} {Analysis} for
  {CBF} {Activation} {Studies} in {Human} {Brain}}.
\newblock \bibinfo{journal}{Journal of Cerebral Blood Flow \& Metabolism}
  \bibinfo{volume}{12}, \bibinfo{pages}{900--918}.
\newblock \bibinfo{note}{Publisher: SAGE Publications Ltd STM}.
\bibitem[{Worsley et~al.(1996)Worsley, Marrett, Neelin, Vandal, Friston and
  Evans}]{worsley_unified_1996}
\bibinfo{author}{Worsley, K.J.}, \bibinfo{author}{Marrett, S.},
  \bibinfo{author}{Neelin, P.}, \bibinfo{author}{Vandal, A.C.},
  \bibinfo{author}{Friston, K.J.}, \bibinfo{author}{Evans, A.C.},
  \bibinfo{year}{1996}.
\newblock \bibinfo{title}{A unified statistical approach for determining
  significant signals in images of cerebral activation}.
\newblock \bibinfo{journal}{Human Brain Mapping} \bibinfo{volume}{4},
  \bibinfo{pages}{58--73}.
\newblock \bibinfo{note}{\_eprint:
  https://onlinelibrary.wiley.com/doi/pdf/10.1002/\%28SICI\%291097-0193\%281996\%294\%3A1\%3C58\%3A\%3AAID-HBM4\%3E3.0.CO\%3B2-O}.
\bibitem[{Yan et~al.(2013)Yan, Cheung, Kelly, Colcombe, Craddock, Di~Martino,
  Li, Zuo, Castellanos and Milham}]{yan_comprehensive_2013}
\bibinfo{author}{Yan, C.G.}, \bibinfo{author}{Cheung, B.},
  \bibinfo{author}{Kelly, C.}, \bibinfo{author}{Colcombe, S.},
  \bibinfo{author}{Craddock, R.C.}, \bibinfo{author}{Di~Martino, A.},
  \bibinfo{author}{Li, Q.}, \bibinfo{author}{Zuo, X.N.},
  \bibinfo{author}{Castellanos, F.X.}, \bibinfo{author}{Milham, M.P.},
  \bibinfo{year}{2013}.
\newblock \bibinfo{title}{A comprehensive assessment of regional variation in
  the impact of head micromovements on functional connectomics}.
\newblock \bibinfo{journal}{NeuroImage} \bibinfo{volume}{76},
  \bibinfo{pages}{183--201}.
\bibitem[{Ye et~al.(2011)Ye, Lazar and Li}]{ye.2011.1}
\bibinfo{author}{Ye, J.}, \bibinfo{author}{Lazar, N.}, \bibinfo{author}{Li,
  Y.}, \bibinfo{year}{2011}.
\newblock \bibinfo{title}{Sparse geostatistical analysis in clustering {fMRI}
  time series}.
\newblock \bibinfo{journal}{Journal of Neuroscience Methods}
  \bibinfo{volume}{199}, \bibinfo{pages}{336--345}.
\bibitem[{Ye et~al.(2009)Ye, Lazar and Li}]{ye_geostatistical_2009}
\bibinfo{author}{Ye, J.}, \bibinfo{author}{Lazar, N.A.}, \bibinfo{author}{Li,
  Y.}, \bibinfo{year}{2009}.
\newblock \bibinfo{title}{Geostatistical analysis in clustering {fMRI} time
  series}.
\newblock \bibinfo{journal}{Statistics in Medicine} \bibinfo{volume}{28},
  \bibinfo{pages}{2490--2508}.
\bibitem[{Zang et~al.(2004)Zang, Jiang, Lu, He and Tian}]{zang_regional_2004}
\bibinfo{author}{Zang, Y.}, \bibinfo{author}{Jiang, T.}, \bibinfo{author}{Lu,
  Y.}, \bibinfo{author}{He, Y.}, \bibinfo{author}{Tian, L.},
  \bibinfo{year}{2004}.
\newblock \bibinfo{title}{Regional homogeneity approach to {fMRI} data
  analysis}.
\newblock \bibinfo{journal}{NeuroImage} \bibinfo{volume}{22},
  \bibinfo{pages}{394--400}.
\bibitem[{Zhang et~al.(2016)Zhang, Cahill, Arbabshirani, White, Baum and
  Michael}]{Zhang2016CA}
\bibinfo{author}{Zhang, C.}, \bibinfo{author}{Cahill, N.},
  \bibinfo{author}{Arbabshirani, M.}, \bibinfo{author}{White, T.},
  \bibinfo{author}{Baum, S.}, \bibinfo{author}{Michael, A.},
  \bibinfo{year}{2016}.
\newblock \bibinfo{title}{Sex and age effects of functional connectivity in
  early adulthood}.
\newblock \bibinfo{journal}{Brain Connectivity} \bibinfo{volume}{6}.

\end{thebibliography}

\appendix

\section{Brain functional connectivity review}

The literature review was conducted on PubMed using the keywords "brain connectivity graph
resting state 'human connectome project'" on September 30, 2021. The search returned 32 papers written between 2014 and 2021. Out of those papers, 5 were not open access and 2 papers were literature reviews, and were not conisdered further. 3 papers were either using seed-based or voxel-to-voxel correlation. Out of the remaining 24 papers 
71\% (17/24) first averaged voxels before computing the inter-regional correlations and 88\% (21/24) employed some kind of spatial aggregation method, including but not limited to averaging over voxels, ICA or dictionary learning.
	
\section{Hypotheses for the spatio-temporal model}
\label{appendix.hypothesis}

The assumptions on the model can be written as follows. For any $i,i^\prime \in \cer$ and $s,t=1,\ldots,T$,
	\begin{align*} 
		\E [X_i(t)] &= \E[\varepsilon_i(t)] = \E[e(t)] = 0, \\
		\E [X_i(s) X_i(t)] &= \E[ \varepsilon_i(s) \varepsilon_i(t)] = E[e(s)e(t)] = 0, \\
		\E [X_i(s) \varepsilon_{i^\prime}(t)] &= \E[X_i(s)e(t)]= \E[ \varepsilon_i(s) e(t)] =  0, \\
		\E [ e(t)^2] &= \sigma^2_e.
	\end{align*}
	
Let $\boldsymbol{\Sigma}$ be the covariance matrix of  the vector $(Y_i(t))_{i\in \mathcal C, t=1,\dots,T}$. In this paper, we assume without referring specifically to this assumption that the parameters $\sigma^2_j$, $\sigma_\varepsilon^2$, $\sigma^2_e$, $\rho_{i i^\prime}$, $\eta_{i i^\prime}$, $r_{jj^\prime}$ are such that $\boldsymbol{\Sigma}$ is a positive definite matrix.

		We also assume that the random variables are independent in time. This is not overly restrictive: in particular, if the random variables have long memory, after a wavelet decomposition, the random variables can be approximated to be decorrelated in time for large wavelet scales \citep{Moulines07SpectralDensity}. In addition, assuming that the $X_i$'s are centered is coherent as it is a well-known fact that a wavelet decomposition based on a wavelet mother with $K$ vanishing moments cancels out every polynomial trend with degree $K-1$.

Finally, to apply the law of large numbers, we also assume that all random variables are absolutely integrable, that is $\E[|Z_i(t)|]<\infty$ for $Z=X,\varepsilon,e$, $i\in \cer$ and $t=1,\dots,T$.

\section{Properties of the estimators of interest} \label{app:proposition1}

For any set of indices $E$ with cardinality $\#E$, we let
		\begin{align}
			\bar \rho_{E} = 	\frac1{(\#E)^2} \sum_{i,i^\prime \in E} \rho_{i i^\prime} 
			\qquad \text{and} \qquad
			\bar \eta_{E} = 	\frac1{(\#E)^2} \sum_{i,i^\prime \in E} \eta_{i i^\prime}\HL{.}
			\label{eq:defrhoetaagg2} 
		\end{align}
	
	The results of the paper are based on this proposition: 
		\jf{
	\begin{prop} \label{prop:results} Consider the notation of Section \ref{sec:model} 
	and assumptions described in~\ref{appendix.hypothesis}. Let $j,j^\prime\in\{1,\dots,J\}$.\\
		
			(i) Let $E \subseteq \r_j$, then for any $t=1,\dots,T$
			\begin{align} 
				\var[ \bar X_E(t) ] &=  \sigma^2_j \; \bar \rho_E  \label{eq:varXbarE} \\
				\var[ \bar \varepsilon_E(t) ] &=  \sigma^2_\varepsilon \; \bar \eta_E  = \mathcal O(1/(\#E)) \label{eq:epsbarE} \\
				\var[ \bar e_E(t) ] &=  \var(e(t)) = \sigma^2_e  \label{eq:varebarE} \\
				\var[ \bar Y_E(t) ] &=  \sigma^2_j \; \bar \rho_E + \sigma^2_\varepsilon \; \bar \eta_E  + \sigma^2_e. \label{eq:varYbarE_appendix} 
			\end{align}
			(ii) Let $E\subseteq \r_j$ and $E^\prime\subseteq \r_{j^\prime}$, then 
			\begin{equation}
				\cov[\bar Y_E(t) , \bar Y_{E^\prime}(t)] =	\left\{
				\begin{array}{ll}
					\sigma_j \sigma_{j^\prime} r_{jj^\prime} + \sigma^2_e & \mbox{ if } j\neq j^\prime \\
					\sigma_j^2 \bar \rho_{E,E^\prime} + \sigma^2_e & \mbox{ if } j=j^\prime
				\end{array}
				\right.
				\label{eq:covYbarEYbarEprime}
			\end{equation} 
			where 
			\begin{equation}
				\bar \rho_{E,E^\prime} = \frac{1}{(\#E)(\#E^\prime)} \sum_{i\in E,i^\prime \in E^\prime} \rho_{|i-i^\prime|}.	
				\label{eq:rhoEEprime}	
			\end{equation}
			(iii) Let $i\in E\subseteq \r_j$ and $i^\prime \in E'\subseteq \r_{j^\prime}$ and assume for any  $i\in E$ and $i^\prime \in E^\prime$
			$|i - i^\prime| \ge p$ (in the case $j=j^\prime$).
			Then as $T \to \infty$, the following statements hold almost surely. 
			\begin{align}
				\widehat \sigma^2( \Vect{Y}_i) &\stackrel{a.s.}{\to} \var[Y_i(1)] 
				\qquad \text{and} \qquad
				\widehat \cov[ \Vect{Y}_i,\Vect{Y}_{i^\prime} ] \stackrel{a.s.}{\to} \cov[Y_i(1),Y_{i^\prime}(1)] \label{eq:convVarCovi} \\ 
				\widehat \sigma^2( \bar{\Vect{Y}}_E) &\stackrel{a.s.}{\to} \var[\bar{Y}_E(1)] 
				\qquad \text{and} \qquad
				\widehat \cov[ \bar{\Vect{Y}}_E,\bar{\Vect{Y}}_{E^\prime} ] \stackrel{a.s.}{\to} \cov[\bar{Y}_E(1),\bar{Y}_{E^\prime}(1)]. \label{eq:convVarCovE} 
			\end{align}

	\end{prop}
	}
	
	\jf{Proposition~\ref{prop:results} is given without proof. (i)-(ii) ensue from the model \eqref{eq:model} while (iii) is quite straightfoward since we have assumed independence in time. 
		
	 As seen from Proposition~\ref{prop:results}, the quantity $\bar \eta_E$ is related to the correlation structure of the local noise. By assuming this noise to be $p$-dependent (that is $\eta_\delta=0$ when $\delta \ge p$), it is clear that the larger $\#E$ the smaller $\bar \eta_E$.}
	
\section{Consistency results for the existing estimators} 

\subsection{Consistency of ${\widehat{r}}_{jj^\prime}^{\CA}$}

	Proposition~\ref{prop:results} shows $\widehat{r}_{jj^\prime}^{\CA}$ is a strongly consistent estimator of $r_{jj^\prime}^\CA$ as $T\to \infty$ where 
		\begin{equation}\label{eq:rCA}
			r^\CA_{jj^\prime}= r_{jj^\prime} \;  \;
			\frac{1 + \sigma_{e,jj^\prime}^2/r_{jj^\prime} }
			{
				\sqrt{
					( \bar \rho_{\r_j} +   \sigma_{\varepsilon,j}^2 \bar \eta_{\r_j}+\sigma_{e,j}^2) 
					( \bar \rho_{\r_{j^\prime}} +  \sigma_{\varepsilon,j^\prime}^2 \bar \eta_{\r_{j^\prime}} +\sigma_{e,j^\prime}^2)
				}
			}.
		\end{equation}
		
			Another way to correct the size effect is to compensate the inter-correlation by the intra-correlation. This \jf{would lead} to the following estimator:
	\begin{equation}
		\widehat{r}_{jj^\prime}^{\widetilde{\AC}}=\frac{1}{N_jN_{j'}}\left (\sum_{i,i' \in \r_j}\widehat{\cor}(\Vect{Y}_i,\Vect{Y}_{i'})\sum_{i,i' \in \r_{j'}}\widehat{\cor}(\Vect{Y}_i,\Vect{Y}_{i'})\right )^{1/2}\widehat{r}^\AC.
		\label{eq.tagg}
	\end{equation}
	The two estimators (\ref{eq.ave}) and (\ref{eq.tagg}) have the important property to remove the size effect (since when $\sigma_\varepsilon=\sigma_e=0$, $r_{jj^\prime}^\AC=r_{jj^\prime}$). Note that both estimators tend to the same limit.

\subsection{Consistency of ${\widehat{r}}_{jj^\prime}^{\AC}$}

Proposition~\ref{prop:results} shows that $\widehat{r}_{jj^\prime}^{\AC}$ is a strongly consistent estimator of $r^\AC_{jj^\prime}$ given by
		\begin{equation}\label{eq:rAC}
			r_{jj^\prime}^\AC = r_{jj^\prime} \;  \; \frac{1+ \sigma_{e,jj^\prime}^2/r_{jj^\prime} }{
				\sqrt{	(1  + \sigma_{\varepsilon,j}^2  +\sigma_{e,j}^2)
					(1 + \sigma_{\varepsilon,j^\prime}^2 +\sigma_{e,j^\prime}^2) }
			}.
	\end{equation}

As revealed by~\eqref{eq:rCA} and~\eqref{eq:rAC}, $\hat r^{\CA}_{jj^\prime}$ and $\hat r^{\AC}_{jj^\prime}$ do not converge toward $r_{jj^\prime}$ when a local noise or global noise is present. We could ask why $\hat r^{\CA}_{jj^\prime}$ is interesting. Actually, a first spatial averaging tends to decrease the effect of the local noise. Indeed, when $\sigma^2_e=0$ (and with equal unit variances to simplify), we have
		\[
		r^\CA_{jj^\prime} = \frac{r_{jj^\prime}}{\sqrt{
				(\bar \rho_{\r_j} + \sigma^2_\varepsilon \bar \eta_{\r_j})
				(\bar \rho_{\r_{j^\prime}} + \sigma^2_\varepsilon \bar \eta_{\r_{j^\prime}} )
		}} \quad \text{ and } \quad
		r^\AC_{jj^\prime} = \frac{r_{jj^\prime}}{1+\sigma_\varepsilon^2}.
		\]
		Hence, if we expect that $\bar \rho_{\r_j}\approx \bar \rho_{\r_{j^\prime}}\approx 1$, $\hat r^\CA_{jj^\prime}$ will be a better estimator since $\bar \eta_{\r_j}=\mathcal O(1/N_j)$. A natural compromise between $\hat r^\CA_{jj^\prime}$ and $\hat r^\CA_{jj^\prime}$ can be defined using local neighborhood as defined by $\lCA$.
	
\subsection{Consistency of ${\widehat{r}}_{jj^\prime}^{\R}$}

From Proposition~\ref{prop:results}, as $T\to \infty$
	\[
		\frac{1}{4}\sum_{\alpha,\beta=1}^2\widehat{\cor}(\Vect{Y}_{i_\alpha^{(b)}},\Vect{Y}_{{i^\prime}_\beta^{(b)}})
		\stackrel{a.s.}{\rightarrow} \frac{\sigma_j \sigma_{j^\prime} r_{jj^\prime}+  \sigma_{e}^2 }{ 
			\sqrt{
				\left(\sigma_j^2 + \sigma_{\varepsilon}^2 + \sigma_{e}^2\right) \left( \sigma_{j^\prime}^2  + \sigma_{\varepsilon}^2 + \sigma_{e}^2\right)}}
		\] 
		and
		\[
		\widehat{\cor}(\Vect{Y}_{i_1^{(b)}} ,\Vect{Y}_{i_2^{(b)}}  )\stackrel{a.s.}{\rightarrow}
		\frac{\sigma_j^2\rho_\delta+ \sigma_{e}^2}{\sigma_j^2+\sigma_{\varepsilon}^2+\sigma_{e}^2},
		\]
	whereby we deduce that $\widehat r^\R$ is a strongly consistent estimator of
		\begin{equation}\label{eq:R}
			r^\R_{jj^\prime} = r_{jj^\prime} \; \; \frac{1 + \sigma_{e,jj^\prime}^2/r_{jj^\prime}}{
				\sqrt{ | (\rho_\delta + \sigma_{e,j}^2)
					(\rho_\delta + \sigma_{e,j^\prime}^2) | } }
			.
		\end{equation}
		From~\eqref{eq:R}, we observe that when $\sigma_e=0$ then for any unknown value of $\sigma_\varepsilon$, $\hat r_{jj^\prime}^\R$ estimates consistently $r_{jj^\prime}/ |\rho_\delta|$ which should be close to $r_{jj^\prime}$ if we take $\delta=p$ and expect that $\rho_p$ is close to 1. In other words, the estimator $\hat r^\R_{jj^\prime}$ is robust to the size of the regions and robust to a possible local noise. 
		
		To reduce the assumption that $\rho_p$ is close to 1, we can combine this idea of replicates with local averaging. This is the topic of the next section.

\section{Consistency of ${\widehat{r}}_{jj^\prime}^{\D}$}

\label{sup:s:prop_proof_D}

The following result is the key ingredient:
		
		\begin{prop}\label{prop:disconnected} Under the notation of this section, as $T \to \infty$, the following statements hold almost surely. \\
			(i) 
			\begin{equation*}
				\widehat \cov ( \Vect{Y}_{i^{(b)}} - \Vect{Y}_{k^{(b)}} , \Vect{Y}_{i^{\prime(b)}} - \Vect{Y}_{k^{\prime(b)}}) \stackrel{a.s.}{\to} \sigma_j \sigma_{j^\prime} r_{jj^\prime}.
			\end{equation*}
			(ii)
			\begin{equation}
				2 \widehat{s}^2(\Vect{Y}_{i^{(b)}} ,\Vect{Y}_{k^{(b)}} , \Vect{Y}_{k^{\prime(b)}} ) \stackrel{a.s.}{\to} 2\sigma_j^2 + 2 \sigma_\varepsilon^2.
			\end{equation}
		\end{prop}

		\begin{proof}
			(i) Using the independence in time, it is clear that the left-hand side converges almost surely to $\cov ( {Y}_{i^{(b)}}(1) - {Y}_{k^{(b)}}(1) , {Y}_{i^{\prime(b)}}(1) - {Y}_{k^{\prime(b)}}(1)) = \sigma_j \sigma_{j^\prime} r_{jj^\prime}+ \sigma_e^2 - 2 \sigma_e^2 + \sigma_e^2$ since the two regions $\r_k$ and $\r_{k^\prime}$ are disconnected,  which leads to the result. \\
			(ii) In the same way, the left-hand side tends to $\var(Y_{i^{(b)}}(1) - {Y}_{k^{(b)}}(1)) + \var(Y_{i^{(b)}}(1) - {Y}_{k^{\prime(b)}}(1)) - \var(Y_{k^{(b)}}(1) - {Y}_{k^{\prime(b)}}(1)) = \sigma_j^2 + \sigma^2_k+\sigma_{k^\prime}^2 + 4 \sigma_\varepsilon^2 - \sigma_k^2-\sigma^2_{k^\prime} - 2\sigma_\varepsilon^2$ which yields the stated limit.
		\end{proof}
		
		In other words, Proposition~\ref{prop:disconnected} shows that $\widehat{r}^\D_{jj^\prime}$ is a strongly consistent estimator of $r^\D_{jj^\prime}$ given by
		\begin{equation}
			r^\D = r_{jj^\prime} \; \; \frac{1 }{
				\sqrt{	\left( 1 +  \sigma_{\varepsilon,j}^2 \right) 
					\left( 1 +  \sigma_{\varepsilon,j^\prime}^2 \right) }
			}
			\label{eq:D}
		\end{equation}
		which, in the situation where $\sigma_\varepsilon=0$, is nothing else than $r_{jj^\prime}$.

\section{Consistency of ${\widehat{r}}_{jj^\prime}^{\RD}$}

\label{sup:s:prop_proof_RD}

The following result is a consequence of~Propositions~\ref{prop:results}-\ref{prop:lD}.
\begin{prop}\label{prop:lrD} As $T \to \infty$, the following statements hold almost surely.\\
(i) For any $i_1,i_2 \in \r_j$, $i_1^\prime,i_2^\prime \in \r_j$ $i_k\in \r_k$ and $i_{k^\prime}\in \r_{k^\prime}$, such that $|i_2-i_1|=|i_2^\prime-i_1^\prime|=\delta\ge p$ 
\begin{equation}
\label{eq:denomRD}
\sqrt{|
\widetilde{\cor}(\Vect{Y}_{i_1},\Vect{Y}_{i_2} ; \Vect{Y}_{i_k},\Vect{Y}_{i_{k^\prime}})
\widetilde{\cor}(\Vect{Y}_{i_1^\prime},\Vect{Y}_{i_2^\prime} ; \Vect{Y}_{i_k},\Vect{Y}_{i_{k^\prime}})
|} \stackrel{a.s.}{\to}
\frac{\sigma_j \sigma_{j^\prime} |\rho_\delta|}{\sqrt{(\sigma_j^2+\sigma_\varepsilon^2)(\sigma_{j^\prime}^2+\sigma_\varepsilon^2)}}
\end{equation}
(ii) For any $\nu$-neighborhoods $\mathcal{V}_{j_1},\mathcal{V}_{j_2}\in \r_j$,
$\mathcal{V}_{j_1^\prime},\mathcal{V}_{j_2^\prime}\in \r_{j^\prime}$
$\mathcal{V}_{k}\in \r_k$
$\mathcal{V}_{k^\prime}\in \r_{k^\prime}$, such that 
for any $i_1 \in \mathcal V_{j_1}$, $i_2 \in \mathcal V_{j_2}, i_1^\prime \in \mathcal V_{j_1}^\prime$, $i_2^\prime \in \mathcal V_{j_2}^\prime $, $| i_1 - i_2 | = | i_1^{\prime} - i_2^{\prime} | = \delta \geq p $
%
\begin{align}
|
\widetilde{\cor}(\bar{\Vect{Y}}_{\mathcal{V}_{j_1}},
\bar{\Vect{Y}}_{\mathcal{V}_{j_2}}; 
\bar{\Vect{Y}}_{\mathcal{V}_{k}},
\bar{\Vect{Y}}_{\mathcal{V}_{k^\prime}})
\widetilde{\cor}(\bar{\Vect{Y}}_{\mathcal{V}_{j_1^\prime}},&
\bar{\Vect{Y}}_{\mathcal{V}_{j_2^\prime}}; 
\bar{\Vect{Y}}_{\mathcal{V}_{k}},
\bar{\Vect{Y}}_{\mathcal{V}_{k^\prime}})
| \stackrel{a.s.}{\to} \nonumber \\
&\frac{\sigma_j^2 \sigma_{j^\prime}^2 \rho^2_{\mathcal V,\mathcal{V}^\prime},\delta}{
(\sigma_j^2\bar \rho_{\mathcal V} +\sigma_{\varepsilon}^2 \bar \eta_{\mathcal V})
(\sigma_{j^\prime}^2\bar \rho_{\mathcal V} +\sigma_{\varepsilon}^2 \bar \eta_{\mathcal V})
} \label{eq:denomlRD}
\end{align}	
where $\mathcal V$, $\mathcal V^\prime$ are two $\nu$-neighborhoods at distance $\delta$.
\end{prop}

Propositions~\ref{prop:lD}-\ref{prop:lrD} show that $\widehat{r}^\RD_{jj^\prime}$ is a strongly consistent estimator of $r^\RD_{jj^\prime}$ given by
\begin{equation}\label{eq:RD}
r^\RD_{jj^\prime} = \frac {r_{jj^\prime}}{|\rho_{\delta}|}
\end{equation}

\section{Consistency of localized versions of estimators}

\subsection{Consistency of ${\widehat{r}}_{jj^\prime}^{\lCA}$}
	
	We can apply Proposition~\ref{prop:results} to show that $r_{jj^\prime}^\lCA$ is a strongly consistent estimator of
		\begin{equation}\label{eq:rlCA}
			r^\lCA_{jj^\prime} = r_{jj^\prime} \;  \;
			\frac{1 + \sigma_{e,jj^\prime}^2/r_{jj^\prime} }
			{
				\sqrt{( \bar \rho_{\mathcal V} +   \sigma_{\varepsilon,j}^2 \bar \eta_{\mathcal V}+\sigma_{e,j}^2)
					( \bar \rho_{{\mathcal V}} + \sigma_{\varepsilon,j^\prime}^2 \bar \eta_{{\mathcal V}} +\sigma_{e,j^\prime}^2)}
			}
	\end{equation}
	\jf{where $\mathcal V$ is any $\nu$-neighborhood. When there is no global noise ($\sigma_e=0$) and for moderate $\nu$, it may be expected than the denominator of $r_{jj^\prime}^\lCA$ is closer to 1 than the ones of $r_{jj^\prime}^\CA$ and $r_{jj^\prime}^\AC$.}
	
\subsection{Consistency of ${\widehat{r}}_{jj^\prime}^{\lR}$}

	Proposition~\ref{prop:results} shows that $\widehat r^\lR_{jj^\prime}$ is a strongly consistent estimator of $r_{jj^\prime}^\lR$ defined by
		\begin{equation} \label{eq:lR}
			r_{jj^\prime}^\lR = 
			r_{jj^\prime} \; \; \frac{1 + \sigma_{e,jj^\prime}^2/r_{jj^\prime}}{
				\sqrt{ | (\bar \rho_{\mathcal V, \mathcal V^\prime,\delta} + \sigma_{e,j}^2)
					(\bar\rho_{\mathcal V, \mathcal V^\prime,\delta} + \sigma_{e,j^\prime}^2) | }} 
		\end{equation}   
		where $\bar\rho_{\mathcal V, \mathcal V^\prime,\delta}$ is defined by~\eqref{eq:rhoEEprime} with $\mathcal V$ and $\mathcal V^\prime$ two $\nu$-neighborhoods at distance~$\delta$. Similarly to the estimator $\hat r^\R$, when $\sigma_e=0$, the previous expression reduces to $r_{jj^\prime}^\lR = 
		r_{jj^\prime}/|\bar\rho_{\mathcal V, \mathcal V^\prime,\delta}|$ and again it is not unreasonable to think that $\bar\rho_{\mathcal V, \mathcal V^\prime,\delta}$ is close to 1.
		
\subsection{Consistency of ${\widehat{r}}_{jj^\prime}^{\lD}$}

	\jf{The following result is an adaptation of Proposition~\ref{prop:disconnected} to local averages.
		\begin{prop}\label{prop:lD} As $T \to \infty$, the following statements hold almost surely. \\
			(i) 
			\begin{equation*}
				\widehat \cov ( \Vect{\bar Y}_{\mathcal V_j^{(b)}} - \Vect{\bar Y}_{\mathcal V_k^{(b)}} , 
				\Vect{\bar Y}_{\mathcal V_{j^\prime}^{(b)}}-\Vect{\bar Y}_{\mathcal V_{k^\prime}^{(b)}}) \stackrel{a.s.}{\to} \sigma_j \sigma_{j^\prime} r_{jj^\prime}.
			\end{equation*}
			(ii)
			\begin{equation}
				2 \widehat{s}^2(\Vect{\bar Y}_{\mathcal V_j^{(b)}},\Vect{\bar Y}_{\mathcal V_k^{(b)}} , \Vect{\bar Y}_{\mathcal V_{k^\prime}^{(b)}} ) \stackrel{a.s.}{\to} 2\sigma_j^2 \bar \rho_{\mathcal V}+ 2 \sigma_\varepsilon^2 \bar \eta_{\mathcal V}.
			\end{equation}
		\end{prop}

		Using Proposition~\ref{prop:lD} (for which proof follows along similar lines as Proposition~\ref{prop:disconnected}), we deduce that $\widehat{r}^\lD_{jj^\prime}$ is a strongly consistent estimator of $r^\lD_{jj^\prime}$ given by}
		\begin{equation} \label{eq:lD}
			r^\lD_{jj^\prime} = r_{jj^\prime} \;\; \frac{1 }{
				\sqrt{(\bar \rho_{\mathcal V} +\sigma_{\varepsilon,j}^2 \bar \eta_{\mathcal V})
					(\bar \rho_{\mathcal V} +\sigma_{\varepsilon,j^\prime}^2 \bar \eta_{\mathcal V})}
			}
		\end{equation}
		where $\mathcal V$ is any $\nu$-neighborhood.

\subsection{Consistency of ${\widehat{r}}_{jj^\prime}^{\lRD}$}

\label{app:consistencylRD}
	Propositions~\ref{prop:lD}-\ref{prop:lrD} show that $\widehat{r}_{jj^\prime}^\lRD$ is a strongly consistent estimator of $r^\lRD_{jj^\prime}$ given by
		\begin{equation}\label{eq:lRD}
			r^\lRD_{jj^\prime} = \frac{r_{jj^\prime}}{|\bar\rho_{\mathcal V, \mathcal V^\prime,\delta}|}
		\end{equation}
		where $\mathcal V$ and $\mathcal V^\prime$ are two $\nu$-neighborhoods at distance $\delta$. Similarly to the previous estimator, $\widehat{r}_{jj^\prime}^\lRD$ is robust to an additive global and local noise.

\end{document}